\documentclass[a4paper,cleveref,USenglish,autoref,thm-restate]{lipics-v2021}

\usepackage{url}
\usepackage{amsfonts}
\usepackage{amsmath}
\usepackage{amssymb}
\usepackage{tikz}
\usetikzlibrary{automata}
\usetikzlibrary{positioning}
\usetikzlibrary{backgrounds}
\usetikzlibrary{external}
\usepackage{wrapfig}
\usepackage{algpseudocodex}
\usepackage{pgfplots}
\usepackage{caption}
\usepackage{subcaption}
\usepackage{pgffor}
\usepackage{todonotes}
\usepackage{verbatim}
\usepackage{svg}

\usepackage{macros}

\pgfplotsset{compat=1.18}
\tikzset{every state/.style={minimum size=2pt}}

\title{Fairness and Consensus in  Opinion Models (Technical Report)}

\ccsdesc{Theory of computation~Social networks}

\titlerunning{Fairness And Consensus in Opinion Transition Systems}  

\author{Jesús Aranda}{Universidad del Valle, Colombia}{jesus.aranda@correounivalle.edu.co}{}{}

\author{Sebastián Betancourt}{Universidad del Valle, Colombia}{joan.betancourt@correounivalle.edu.co}{}{}

\author{Juan Fco. Díaz}{Universidad del Valle, Colombia}{juanfco.diaz@correounivalle.edu.co}{}{}

\author{Frank Valencia}{CNRS LIX, École Polytechnique de Paris, France \and Pontificia Universidad Javeriana Cali, Colombia}{frank.valencia@gmail.com}{}{}

\authorrunning{Aranda, et al.}

\Copyright{Aranda, Betancourt,  Díaz and Valencia}

\keywords{Social networks, fairness, DeGroot, consensus, asynchrony}

\supplementdetails[subcategory={Source python code for simulations},]
{Software}{https://github.com/promueva/Fairness-and-Consensus-in-Opinion-Models}

\funding{This work is supported by the Colombian Minciencias project PROMUEVA, BPIN 2021000100160.}

\EventEditors{John Q. Open and Joan R. Access}
\EventNoEds{2}
\EventLongTitle{42nd Conference on Very Important Topics (CVIT 2016)}
\EventShortTitle{CVIT 2016}
\EventAcronym{CVIT}
\EventYear{2016}
\EventDate{December 24--27, 2016}
\EventLocation{Little Whinging, United Kingdom}
\EventLogo{}
\SeriesVolume{42}
\ArticleNo{23}

\begin{document}

\maketitle              

\begin{abstract}
We introduce a DeGroot-based model for opinion dynamics in social networks. A community of agents is represented as a weighted directed graph whose edges indicate how much agents influence one another. The model is formalized  using labeled transition systems, henceforth called \emph{opinion transition systems (OTS)}, whose states represent the agents' opinions and whose actions are the edges of the influence graph. If a transition labeled $(i,j)$ is performed, agent $j$ updates their opinion taking into account the opinion of agent $i$ and the influence $i$ has over $j$. We study \emph{(convergence to) opinion consensus} among the agents of strongly-connected graphs with influence values in the interval $(0,1)$.  We show that consensus cannot be guaranteed under the standard \emph{strong fairness} assumption on transition systems. We derive that consensus is guaranteed under a stronger notion from the literature of concurrent systems; \emph{bounded fairness}. We argue that bounded-fairness is too strong of a notion for consensus  as it almost surely rules out random runs and it is not a constructive liveness property. We introduce a weaker fairness notion, called \emph{$m$-bounded fairness}, and show that it guarantees consensus. The new notion includes almost surely all random runs and it is a constructive liveness property. Finally, we consider OTS with \emph{dynamic influence} and show convergence to consensus holds under $m$-bounded fairness if the influence changes within a fixed interval $[L,U]$ with $0<L<U<1$. We illustrate OTS with examples and simulations, offering insights into  opinion formation under fairness and dynamic influence.

\end{abstract}%

\section{Introduction}
Social networks have a strong impact on \emph{opinion formation}, often resulting in polarization. Broadly, the dynamics of opinion formation in social networks involve users expressing their opinions, being exposed to the opinions of others, and potentially adapting their own views based on these interactions. Modeling these dynamics enables us to glean insights into how opinions form and spread within social networks.

The DeGroot model \cite{degroot} is one of the most prominent formalisms for social learning and opinion formation dynamics, and it remains a continuous focus of study in social network theory \cite{Golub2017survey}. A given community is represented as a weighted directed graph, known as the \emph{influence graph}, whose edges indicate how much individuals (\emph{agents}) influence one another.  Each agent has an opinion represented as a value in $[0,1]$, indicating the strength of their agreement with an underlying proposition (e.g., ``\emph{AI poses a threat to humanity}''). Agents repetitively revise their opinions by averaging them with those of their contacts, taking into account the influence each contact holds. (There is empirical evidence validating the opinion formation through averaging of the model in controlled sociological experiments, e.g., \cite{degrootEmpirico2015}.) A fundamental theoretical result of the model states that the agents will \emph{converge to consensus} if the influence graph is \emph{strongly connected} and the agents have non-zero self-influence (\emph{puppet freedom}) \cite{Golub2017survey}. The significance of this result lies in the fact that consensus is a central problem in social learning. Indeed, the inability to reach consensus is a sign of a polarized community.

Nevertheless, the DeGroot model makes at least two assumptions that could be overly constraining within social network contexts. Firstly, it assumes that all the agents update their opinions simultaneously (\emph{full synchrony}), and secondly, it assumes that the influence of agents remains the same throughout opinion evolution (\emph{static influence}).  These assumptions may hold in some controlled scenarios and render the model tractable but in many real-world scenarios individuals do not update their opinions simultaneously \cite{Noorazar2020}. Instead, opinion updating often occurs \emph{asynchronously}, with different agents updating their opinions at different times. Furthermore, individuals may gain or lose influence through various factors, such as expressing contrarian or extreme opinions \cite{Goldsmith2015}. 

In this paper, we introduce an \emph{asynchronous} DeGroot-based model with \emph{dynamic influence} to reason about opinion formation, building upon notions from concurrency theory. The model is presented by means of labeled transition systems, here called \emph{opinion transition systems (OTS)}. The states of an OTS represent the agents' opinions, and the actions (labels) are the edges of the influence graph. All actions are \emph{always} enabled.  If a transition labeled with an edge $(i,j)$ is chosen, agent $j$ updates their opinion by averaging it with the opinion of agent $i$ weighted by the influence that this agent carries over $j$. A \emph{run} of an OTS is an infinite sequence of (chosen) transitions.

We shall focus on the problem of convergence to opinion consensus in runs of the OTS, \emph{assuming} strong connectivity of the influence graph and puppet freedom. For consensus to make sense, all agents should have the chance to update their opinions. Therefore, we need to make  \emph{fairness} assumptions about the runs. In concurrency theory, this means requiring that some actions be performed sufficiently often. 

We first show that contrary to the DeGroot model, consensus \emph{cannot} be guaranteed for runs of OTS even under the standard \emph{strong fairness} assumption (i.e., that each action occurs infinitely often in the run) \cite{GrumbergOrna1985,Lehmann1981}. This highlights the impact of asynchronous behavior on opinion formation.

We then consider the well-known notion of \emph{bounded fairness} in the literature on verification of concurrent systems \cite{Dershowitz:03}. This notion requires that every action must be performed not just eventually but within some bounded period of time. We show that bounded-fairness guarantees convergence to consensus. This also gives us insight into opinion formation through averaging, i.e., preventing unbounded  delays of actions (opinion updates) is sufficient for convergence to consensus. 

Nevertheless, bounded fairness does not have some properties one may wish in a fairness notion. In particular, it is not a \emph{constructive liveness} property in the sense of \cite{Varacca2005concur,Varacca2012}. Roughly speaking, a fairness notion is a constructive liveness property if, while it may require that a particular action is taken sufficiently often, it should not prevent any other action from being taken sufficiently often. Indeed, we will show that preventing unbounded delays implies preventing some actions from occurring sufficiently often. 

Furthermore, bounded-fairness is not \emph{random inclusive}. A fairness notion is random inclusive if any random run (i.e., a run where each action is chosen independently  with non-zero probability) is \emph{almost surely} fair under the notion. We find this property relevant because we wish to apply our results to other asynchronous randomized models whose runs are random and whose opinion dynamics can be captured as an OTS.

We therefore introduce a new weaker fairness notion, called \emph{$m$-bounded fairness}, and show that it guarantees consensus. The new notion is shown to be a constructive liveness property and random inclusive. We also show that consensus is guaranteed under $m$-bounded fairness even if we allow for \emph{dynamic influence} as long as all the changes of influence are within a fixed interval $[L,U]$ with $0<L<U<1$.  

All in all, we believe that asynchronous opinion updates and dynamic influence provide us with a model more faithful to reality than the original DeGroot model. The fairness assumptions and consensus results presented in this paper show that the model is also tractable and that it brings new insights into opinion formation in social networks. To the best of our knowledge, this is the first work using tools and concepts from concurrency theory in the context of opinion dynamics and social learning.

Furthermore, since \emph{$m$-bounded fairness} is random inclusive, our result extends with dynamic influence the consensus result in \cite{Fagnani2008} for distributed averaging with randomized gossip algorithms. Distributed averaging is a central problem in other application areas, such as decentralized computation, sensor networks and clock synchronization.

{\bf Organization.} The paper is organized as follows: In Section \ref{sect:Model}, we introduce OTS and the consensus problem. Initially, to isolate the challenges of asynchronous communication in achieving consensus, we assume static influence.  In Section \ref{sect:consensusArgument}, we identify counter-examples, graph conditions, and fairness notions for consensus to give some insight into opinion dynamics. In Section \ref{section:m-fairness}, we introduce a new notion of fairness and state our first consensus theorem. Finally, in Section \ref{sect:dynamicInfluence}, we add dynamic influence and give the second consensus theorem. 

The proofs are included in the appendix. The Python code used to produce OTS examples and simulations in this paper can be provided to the CONCUR PC Chairs upon request.

%

\section{The Model}
\label{sect:Model}

In the standard DeGroot model \cite{degroot}, agents update their opinion \emph{synchronously} in the following sense: at each time unit, all the agents (individuals) update simultaneously their current opinion by listening to the current opinion values of those who influence them. This notion of updating may be unrealistic in some social network scenarios, as individuals may listen to (or read) others' opinions at different points in time. 

In this section, we introduce an opinion model where individuals update their beliefs asynchronously; one agent at a time updates their opinion by listening to the opinion of one of their influencers.

\subsection{Opinion Transition Systems} In social learning models, a \emph{community} is typically represented as a directed weighted graph with edges between individuals (agents) representing the direction and strength of the influence that one has over the other. This graph is referred to as the \emph{Influence Graph}. 

\begin{definition}[Influence Graph]\label{def:influence-graph}
    An  \emph{influence graph} is a directed weighted graph $G=(A,E,I)$ with $A=\{1,\ldots,\mathbf{n}\}$, $\mathbf{n}>1$, the vertices, $E\subseteq A^2 - \mbox{Id}_A$ the edges (where $\mbox{Id}_A$ is the identity relation on $A$) and $I:E \to (0,1]$ the weight function.
\end{definition}

 The vertices in $A$ represent $\mathbf{n}$ agents of a given community or network. The set of edges $E$ represents the (direct) influence relation between agents; i.e., $(i,j)\in E$ means that agent $i$ influences agent $j$. The value $I(i,j)$, for simplicity written $\I{(i,j)}$ or $\I{ij}$ , denotes the strength of the influence: a higher value means stronger influence.

Similar to the DeGroot-like models in \cite{Golub2017survey}, we model the evolution of  agents' opinions about some underlying \emph{statement} or \emph{proposition}, such as, for example,  ``\emph{human activity has little impact on climate change}'' or ``\emph{AI poses a threat to humanity}''.

The \emph{state of opinion} (or \emph{belief state}) of  all the agents is represented as a \emph{vector} in $[0,1]^{|A|}$. If $\B{}{}$ is a state of opinion, $\B{}{i}$ denotes the \emph{opinion} (\emph{belief}, or \emph{agreement}) value of agent $i \in A$ regarding the underlying proposition: the higher the value of $\B{}{i}$, the stronger the agreement with such a proposition. If $\B{}{i}=0$, agent $i$ completely \emph{disagrees} with the underlying proposition; if $\B{}{i}=1$, agent $i$ completely \emph{agrees} with the underlying proposition.  

The opinion state is updated as follows: Starting from an initial state, at each time unit, one of the agents, say $j$, updates their opinion  taking into account  the influence and the opinion of one of their contacts, say $i$. Intuitively, in social network scenarios, this can be thought of as having an agent $j$ read or listen to the opinion of one of their influencers $i$ and adjusting their opinion $\B{}{j}$ accordingly. 

The above intuition can be realized as  a \emph{Labelled Transition System} (LTS) whose set of states is $S=[0,1]^{|A|}$ and set of \emph{actions} is $E$.  

\begin{definition}[OTS]\label{def:ots}
    An Opinion Transition System (OTS) is a tuple $M=(\graph,\binit,\to)$ where $\graph=(\agents,\edges,I)$ is an influence graph, $\binit \in S=[0,1]^{|A|}$ is the  initial opinion state, and 
    $\to \subseteq S \times \edges\times S$ is a \textit{(labelled)  transition relation} defined thus: $(\B{}{},(i,j),\B[B']{}{})\in \to$, written $\B{}{} \xrightarrow{(i,j)} \B[B']{}{}$, \text{ iff for every}\  $k\in \agents,$ 
            \begin{equation}\label{eq:reduction}
            \B[B']{}{k} = 
\begin{cases} 
    \B{}{j} + (\B{}{i} - \B{}{j})\I{ij} & \text{if } k=j \\
    \B{}{k} & \text{otherwise}
\end{cases} \end{equation}
If $\B{}{} \xrightarrow{e} \B[B']{}{}$ we say that $\B{}{}$ evolves into   $\B[B']{}{}$ by performing (choosing or executing) the \emph{action} $e$. 
\end{definition}

A labeled transition $\B{}{} \xrightarrow{(i,j)} \B[B']{}{}$ represents the opinion evolution from $\B{}{}$ to $\B[B']{}{}$ when choosing an action represented by the edge $(i,j)$. As a result of this action, agent $j$ updates their opinion as $\B{}{j} + (\B{}{i} - \B{}{j})\I{ij}$, thereby moving closer to the opinion of agent $i$. Alternatively, think of agent $i$ as pulling the opinion of agent $j$ towards $\B{}{i}$. The  higher the influence of $i$ over $j$, $\I{ij}$, the closer it gets. 
Intuitively, if $\I{ij}<1$, it means that agent $j$ is \emph{receptive} to agent $i$ but offers certain \emph{resistance} to fully adopting their opinion.  If $\I{ij}=1$, agent $j$ may be viewed as a \emph{puppet} of $i$ who disregards (or forgets) their own opinion to adopt that of $i$.

\begin{remark}
    In Def. \ref{def:influence-graph}, we do not allow edges of the form $(j,j)$. In fact, allowing them would \emph{not} present us with any additional technical issues, and the results in this paper would still hold. The reason for this design choice, however, has to do with clarity about the intended intuitive meaning of a transition. Suppose that  $\B{}{} \xrightarrow{(i,j)} \B[B']{}{}$. Since $\B[B']{}{j}=\B{}{j} + (\B{}{i} - \B{}{j})\I{ij} = \B{}{j}(1-\I{ij}) + \B{}{i}\I{ij}$, agent $j$ gives a weight of $\I{ij}$ to the opinion of $i$ and of $(1-\I{ij})$ to \emph{their own opinion}. Therefore, the weight that $j$ gives to their opinion may change depending on the agent $i$. Thus, allowing also a fixed weight $\I{jj}$ of agent $j$ to their own opinion may seem somewhat confusing to some readers. Furthermore, for any $\B{}{}\in S$ we would have  $\B{}{} \xrightarrow{(j,j)} \B{}{}$  regardless of the value $I_{jj}$ thus making the actual value irrelevant. Notice also we do not require the sum of the influences over an agent to be 1.
\end{remark}

\subsection{Runs and Consensus}

We are interested in properties of opinion systems, such as convergence to consensus and fairness, which are inherent properties of infinite runs of these systems. 

\begin{definition}[e-path, runs and words]\label{def:runs}
 An \emph{execution path (e-path)} of an OTS  $M=(\graph,\binit,\to)$, where $\graph=(\agents,\edges,I)$,  is an infinite  sequence  $\pi = \B{0}{}e_0\B{1}{}e_1\ldots $ (also written $\execution$) such that $\B{t}{} \xrightarrow{e_t} \B{t+1}{}$ for each $t\in\Nat$. We say that $e_t$ is the action performed at time $t$ and that $B_t$ is the state of opinion at time $t$.  Furthermore, if $\B{0}{} = \binit$ then the e-path $\pi$ is said to be a \emph{run} of $M$. 
 
An $\omega$-\emph{word} of $M$ is an infinite sequence of edges (i.e, an element of $E^\omega$). The sequence $w_\pi= \word$ is the $\omega$-word generated by $\pi.$ Conversely, given an $\omega$-word $w=e'_0.e'_1\ldots$ the (unique) run that corresponds to it is $\pi_w = \binit\xrightarrow{e'_0}\B{1}{}\xrightarrow{e'_1}{}\ldots$


\end{definition}

\begin{remark}
The uniqueness of the run that corresponds to a given $\omega$-word is derived from the fact that an OTS is a deterministic transition system. This gives us a one-to-one correspondence between $\omega$-words and runs, which allows us to abstract away from opinion states when they are irrelevant or clear from the context. In fact, throughout the paper, \emph{we will use the terms $\omega$-words and runs of an OTS interchangeably when no confusion arises}. 
It is also worth noting that in OTS, any action (edge) can be chosen at any point in an execution path; that is, \emph{all actions are enabled}. 
\end{remark}

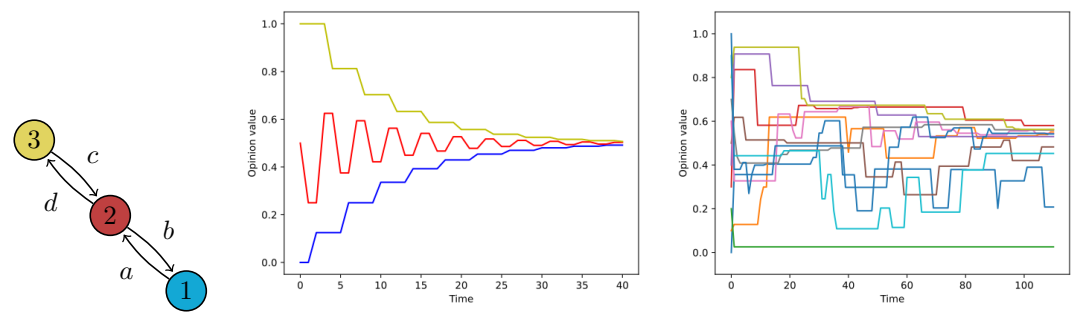
\begin{figure}[t]
\centering%
\begin{subfigure}[t]{0.19\textwidth}%
\centering

    \begin{tikzpicture}[->, 
    shorten >=1pt,
    semithick,
    every state/.style={inner sep=0pt, minimum size=15pt},
    bend angle=45, 
    initial text=,]
    \node[state, fill=yellow!70!gray] (3) at (0,2) {3};
    \node[state, fill=red!50!gray] (2) at (1,1) {2};
    \node[state, fill=cyan!70!gray] (1) at (2,0) {1};

    \path(1) edge[right, anchor=north east,  bend left=10] node[align=center]{$a$}
    (2);

    \path(2) edge[left, anchor=south west ,  bend left=10] node[align=center]{$b$}
    (1);

    \path(3) edge[right, anchor=south west,  bend left=10] node[align=left]{$c$}
    (2);

    \path(2) edge[left, anchor=north east,  bend left=10] node[align=right]{$d$}
    (3);

    \end{tikzpicture}
\caption{\label{fig:ex1}  OTS with influence graph with agents $A=\{1,2,3\}$, edges $E=\{a,b,c,d\}$, influence $I_e = 1/2$ for all $e\in E$ and $\binit=(0,0.5,1)$.}
\end{subfigure}
\hfill
\begin{subfigure}[t]{0.39\linewidth}
\centering%
\includegraphics[width=1.1\textwidth]{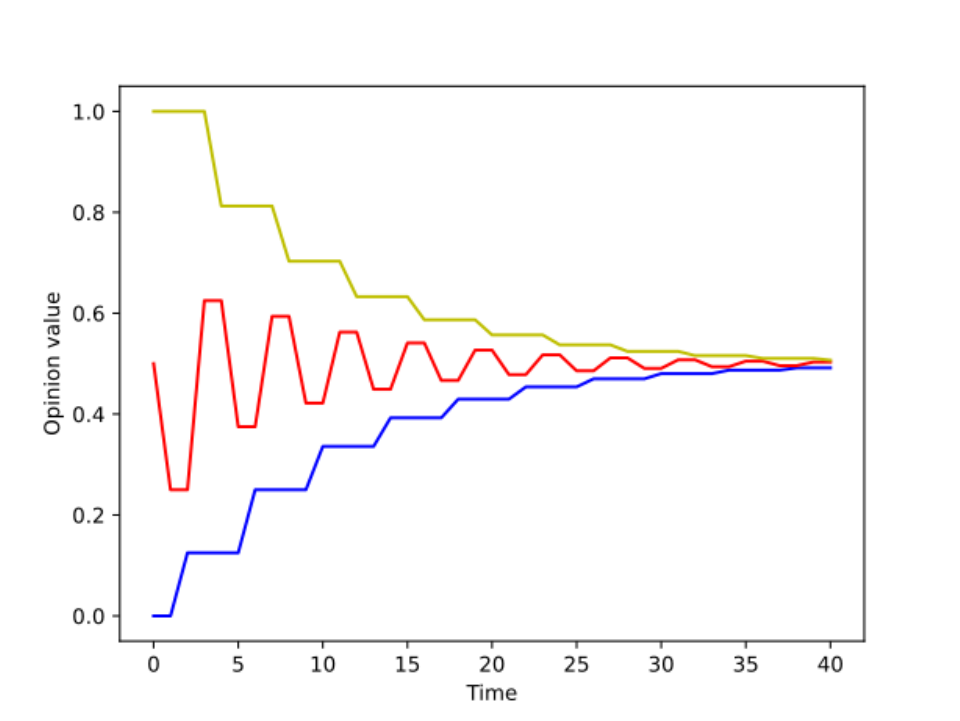}
\caption{\label{fig:consensusExample} Opinion evolution for the run that corresponds to $(abcd)^\omega$ of the OTS in Fig. \ref{fig:ex1}. Each plot corresponds to the opinion evolution of the agent with the same color.}
\end{subfigure}
\hfill
\begin{subfigure}[t]{0.39\textwidth}%
    \centering
    \includegraphics[width=1.1\textwidth]{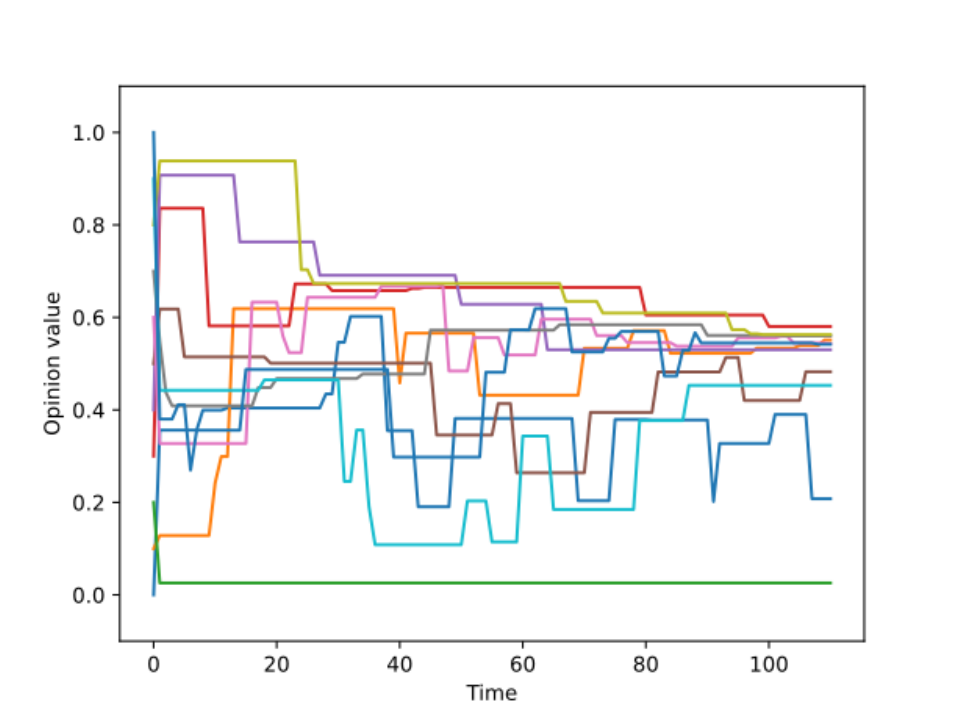}
\caption{Opinion evolution of a run of an OTS with a $\graph=(\agents,\edges,\I{})$, $\agents=\{1,\ldots,11\}$,  $I_e= 0.5$ for each $e \in \edges$, $\binit=(0, 0.1, 0.2, 0.3, 0.4, 0.5, 0.6, 0.7, 0.8, 0.9, 1.0)$. Each edge of $G$ was generated with prob. $0.3$. The edges of the (partial) run were uniformly chosen from $E$.
    \label{fig:introEx3}}
\end{subfigure}

\caption{\label{fig:intro}Run examples for OTS in Fig. \ref{fig:ex1} and randomly-generated OTS in 
Fig. \ref{fig:introEx3}}.

\end{figure}

Consensus is a property of central interest in social learning models \cite{Golub2017survey}. Indeed, failure to reach a consensus is often an indicator of  polarization in a community. 

\begin{definition}[Consensus]\label{def:consensus}
 Let  $M=(\graph,\binit,\to)$ be an OTS with $\graphElem$ and $\pi=\run$ be a run. We say that an agent $i\in A$ \emph{converges} to an opinion value $v\in[0,1]$ in  $\pi$ if   $\lim_{t \to \infty} \B{t}{i} = v$. The run $\pi$ \emph{converges to consensus} if all the agents in $A$ converge to the \emph{same} opinion value in $\pi$. 

 Furthermore,  $\B{}{}$ is said to be a \emph{consensual state} if it is a constant vector; i.e., if there exists $v\in[0,1]$ such 
 that for every $i\in A$, $\B{}{i} = v$. 
\end{definition}

\begin{example}\label{ex:sec:model} Let $M=(\graph,\binit,\to)$ where $\graph$ is the influence graph in Fig. \ref{fig:ex1} and $\binit=(0,0.5,1)$. If we perform  $a$ on $\binit$ we obtain $\binit\xrightarrow{a}\B{1}{}=(0.0,0.25,1.0)$. 

Consider the word $w=(abcd)^\omega$. Then  $\pi_w=\binit\xrightarrow{a}
(0.0,0.25,1.0)\xrightarrow{b}
(0.125,0.25,1.0)\xrightarrow{c}(0.125,0.625,1.0)\xrightarrow{d}(0.125,0.625,0.8125)\xrightarrow{a}\ldots$. Fig. \ref{fig:consensusExample} suggests that $\pi_w$ indeed converges to consensus (to opinion value 0.5). 
 A more complex opinion evolution example from a randomly generated graph with eleven agents is illustrated in Fig. \ref{fig:introEx3}. 
\end{example}

The examples above illustrate runs that may or may not converge to consensus.
In the next section, we identify conditions on the influence and topology of graphs and on the runs that guarantee this central property of opinion models.%

\section{Strong Connectivity, Puppet-Freedom and Fairness }\label{sect:consensusArgument}

In this section, we discuss graph properties, as well as fairness notions and criteria from the literature on concurrent systems that give us insight into how agents converge to consensus in an OTS. For simplicity, we assume an underlying OTS $M=(\graph,\binit,\to)$ with an influence graph $\graph=(\agents,\edges,I)$. We presuppose basic knowledge of graph theory and formal languages.

\subsection{Strong Connectivity}
\label{sec:strong-connectivity}

\begin{figure}[t]
    \centering%
    \begin{subfigure}[t]{0.19\textwidth}%
    \centering
    
        \begin{tikzpicture}[->, 
        shorten >=1pt, 
        node distance = 1cm,
        semithick,        
        bend angle=45, 
        initial text=,]
        \node[state, fill=blue!60!white] (0) at (0,0) {1};
        \node[state, fill=orange] (1) at (1.5,0) {2};
        \node[state, fill=green!50!gray] (2) at (0,1.5) {3};
        \node[state, fill=red!50!gray] (3) at (1.5,1.5) {4};
        \node[state, fill=magenta!50!gray] (4) at (0,3) {5};
        \node[state, fill=purple!50!gray] (5) at (1.5,3) {6};

        \path(1) edge[right, anchor=north,  bend left=10] node[align=center]{$a_1$}
        (0);
        \path(0) edge[right, anchor=south,  bend left=10] node[align=center]{$a_0$}
        (1);

        \path(2) edge[right, anchor=south,  bend left=10] node[align=center]{$a_2$}
        (3);
        \path(3) edge[right, anchor=north,  bend left=10] node[align=center]{$a_3$}
        (2);

        \path(4) edge[right, anchor=south,  bend left=10] node[align=center]{$a_4$}
        (5);
        \path(5) edge[right, anchor=north,  bend left=10] node[align=center]{$a_5$}
        (4);
        
        \path(4) edge[right, anchor=east] node[align=center]{$a_6$}
        (2);
        \path(5) edge[right, anchor=west] node[align=center]{$a_7$}
        (3);

        \path(0) edge[right, anchor=east] node[align=center]{$a_8$}
        (2);
        \path(1) edge[right, anchor=west] node[align=center]{$a_9$}
        (3);
       
        \end{tikzpicture}
    \caption{\label{fig:strConnCounterexGraph} Influence graph with $I_e = 1/2$ for all $e\in E$ and $\binit=(0.4,0.5,0.45, 0.55,$ $ 0.5,0.6)$} 
    \end{subfigure}
    \hfill
    \begin{subfigure}[t]{0.39\linewidth}
    \centering%
    \includegraphics[width=1.1\textwidth]{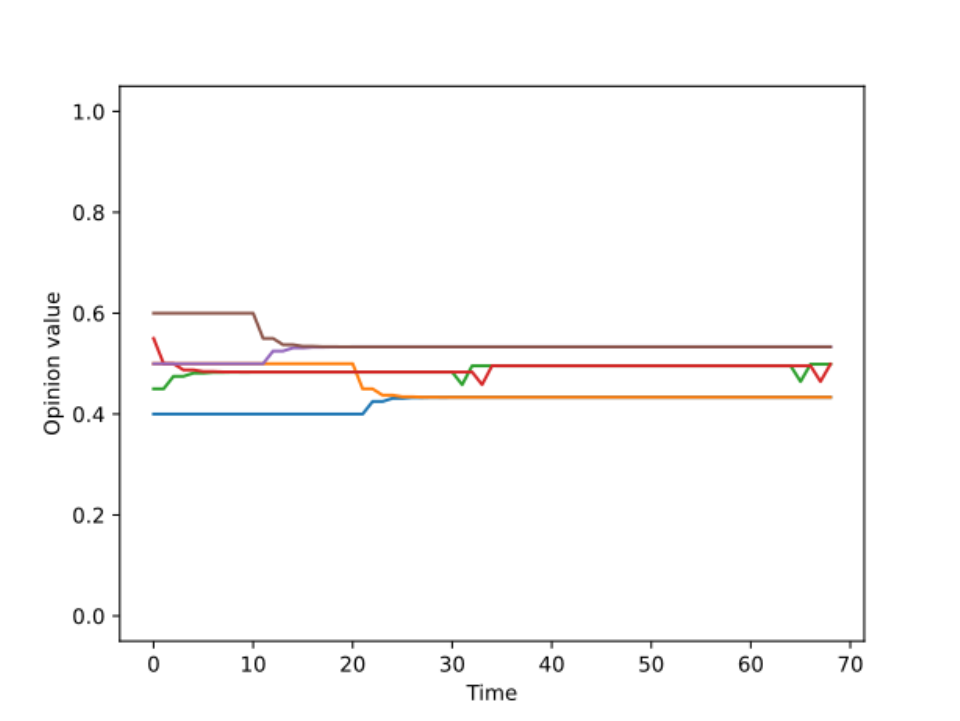}
    \caption{\label{fig:strConnCounterexPlot} Opinion evolution of the run 
    $((a_2a_3)^5(a_4a_5)^5(a_0a_1)^5(a_6a_7a_8a_9))^\omega$. Each plot corresponds to the opinion evolution of the agent with the same color in  Fig. \ref{fig:strConnCounterexGraph}.}
    \end{subfigure}
    \hfill
    \begin{subfigure}[t]{0.39\linewidth}
        \centering%
    \includegraphics[width=1.1\textwidth]{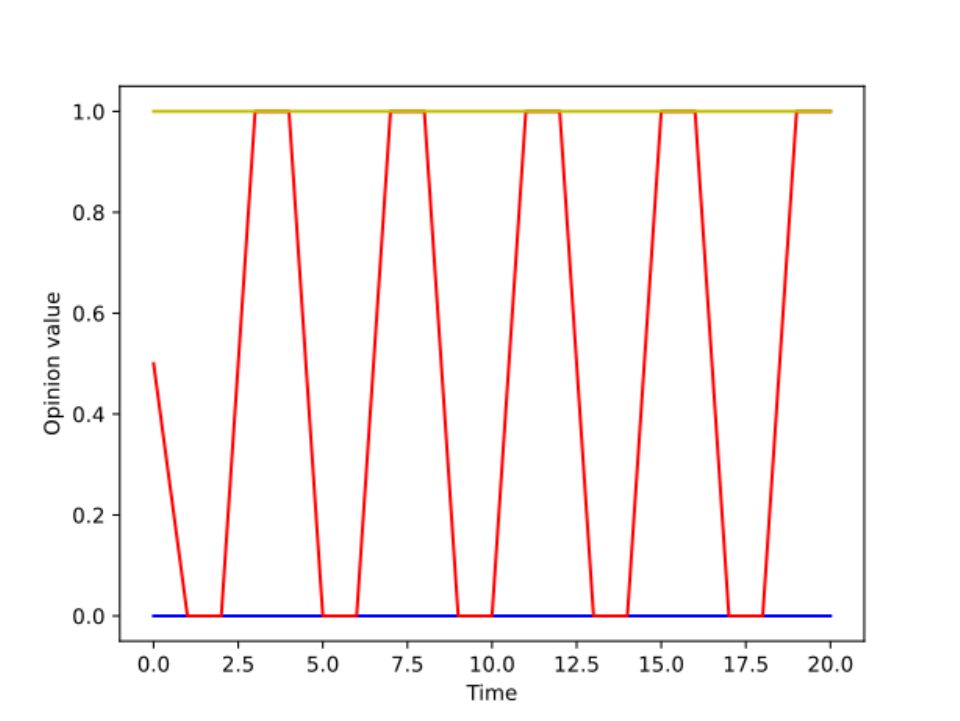}
\caption{\label{fig:puppets} Opinion evolution of the run $(abcd)^\omega$ for an OTS with $\graph$ and $\binit$ from Fig. \ref{fig:ex1} but assuming $I_e=1$ for all $e \in \edges$. }
    \end{subfigure}

    \caption{Run examples for OTS in Fig. \ref{fig:ex1} and Fig. \ref{fig:strConnCounterexGraph}}
    \label{fig:strConnCounterex}

    \end{figure}
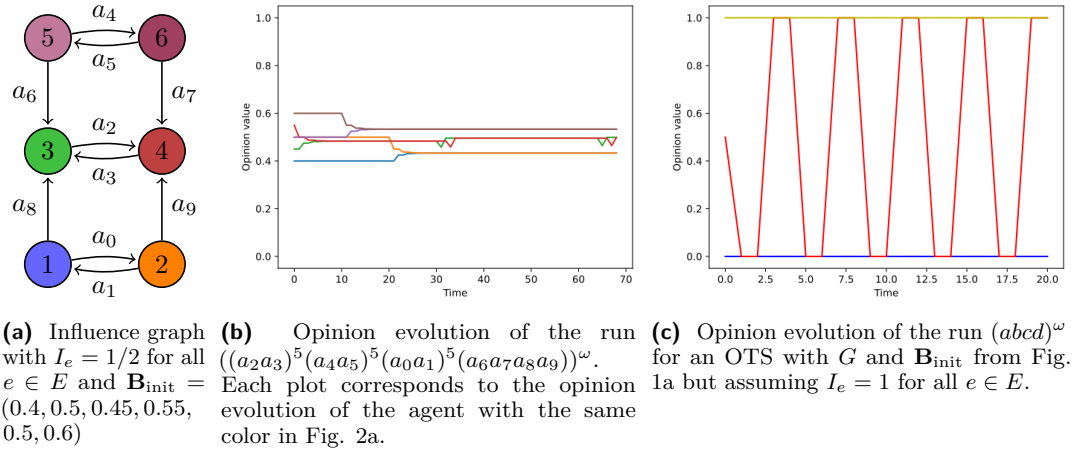

As in the DeGroot model, if there are (groups of) agents in $G$ that do not influence each other (directly or indirectly) and their initial opinions are different, these groups may converge to different opinion values. Consider the example in Fig. \ref{fig:strConnCounterex} where the groups of agents $G_1=\{1,2\}$ and $G_2=\{ 5,6\}$ do not have external influence (directly or indirectly), but influence the group $G_3=\{3,4\}$. Each group is strongly connected within;  their members influence each other. The agents in $G_1$ converge to an opinion, and so do the agents in $G_2$, but to a different one. Hence, the agents in both groups cannot converge to consensus.  The agents in $G_3$ do not even converge to an opinion because they are regularly influenced by the dissenting opinions of $G_1$ and $G_2$.

The above can be prevented by requiring \emph{strong connectivity}, i.e., there must be a path in $G$ from any other to any other. Recall that a \emph{graph path} from $i$ to $j$ of length $m$ in $G$ is a sequence of edges of $E$ of the form $(i,i_1)(i_1,i_2)\ldots(i_{m-1},j)$, where the agents in the sequence are distinct. We shall refer to graph paths as \emph{g-paths} to distinguish them from e-paths in Def. \ref{def:runs}. We say that agent $i$ influences agent $j$ if there is a g-path from $i$ to $j$ in $G$. The graph $G$ is \emph{strongly connected} iff there is a g-path from any agent to any other in $G$. Hence, in strongly-connected graphs, all agents influence one another.

\subsection{Puppet-Freedom} 
\label{sect:puppetFreedom}

Nevertheless, too much influence may prevent consensus. If $\B{}{} \xrightarrow{(i,j)} \B[B']{}{}$ and $\I{ij}=1$, agent $j$ behaves as a \emph{puppet} of $i$ forgetting their own opinion and adopting that of $j$.  Fig. \ref{fig:puppets}  illustrates this for the strongly-connected graph in Fig. \ref{fig:ex1} but with $I_{ij}=1$ for each $(i,j)\in E$: Agents 1 and 3 use Agent 2 as a puppet, constantly swaying his opinion between 0 and 1. We therefore say that the influence graph $G$ is \emph{puppet free} if for each $(i,j)\in E$, $I_{ij}<1.$

\subsection{Strong Fairness}

In an OTS, if $G$ is strongly connected but a given edge is never chosen in a run (or not chosen sufficiently often), it may amount to not having all agents influence each other in that run, hence preventing consensus. For this reason, we make some fairness assumptions about the runs.

In the realm of transition systems, fairness assumptions rule out some runs, typically those where some actions are not chosen sufficiently often when they are enabled sufficiently often. There are many notions of fairness (see \cite{Apt1987feasibility, Glabbeek, Kwiatkowska1989survey} for surveys), but strong fairness is perhaps one of the most representative. As noted above, every action $e \in E$ is always enabled in every run of an OTS. Thus, in our context, strong fairness of a given OTS run ($\omega$-word) amounts to requiring that every action $e$ occurs infinitely often in the run. 

\begin{definition}[Strong fairness]\label{def:strongfairness} Let $w$ be an $\omega$-word of an OTS. We say that $w$ is \emph{strongly fair} if every $e\in E$ occurs in every suffix of $w$.
\end{definition}

\begin{figure}[t]

\centering%
\begin{subfigure}[t]{0.49\textwidth}%
    \centering%
    \includegraphics[width=0.8\textwidth]{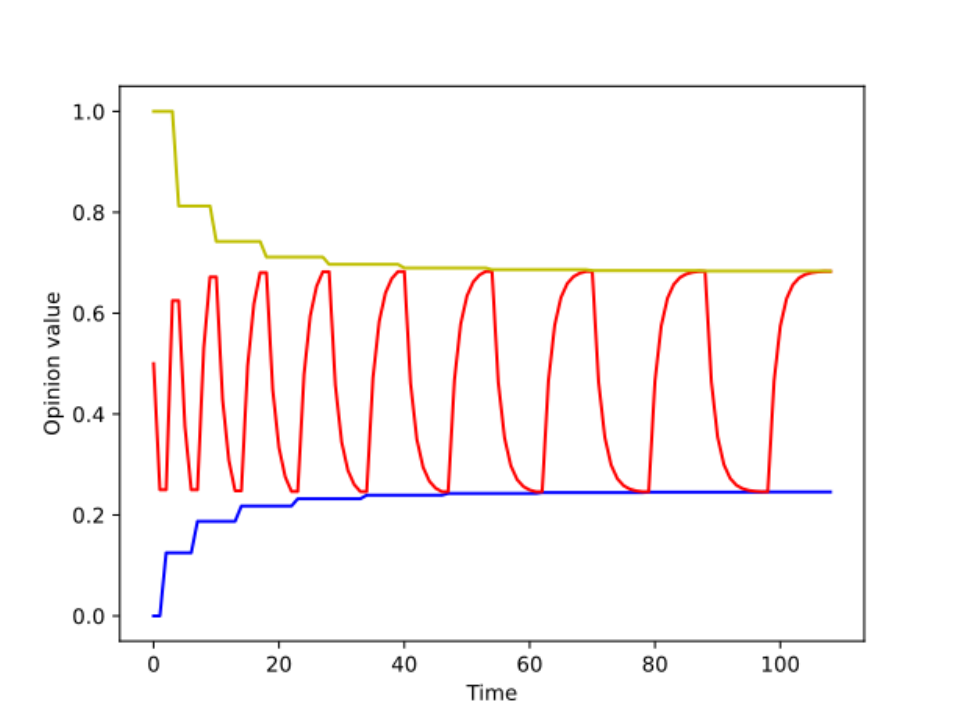}
\caption{\label{fig:edgeFairnessCounterex1}Opinion evolution of the OTS from Fig. \ref{fig:ex1} for the
$\omega$-word $u=(a^{n}bc^{n}d)_{n\in\Nat^+}$}

\end{subfigure}
\hfill
    \begin{subfigure}[t]{0.49\linewidth}
    \centering%
    \includegraphics[width=0.8\textwidth]{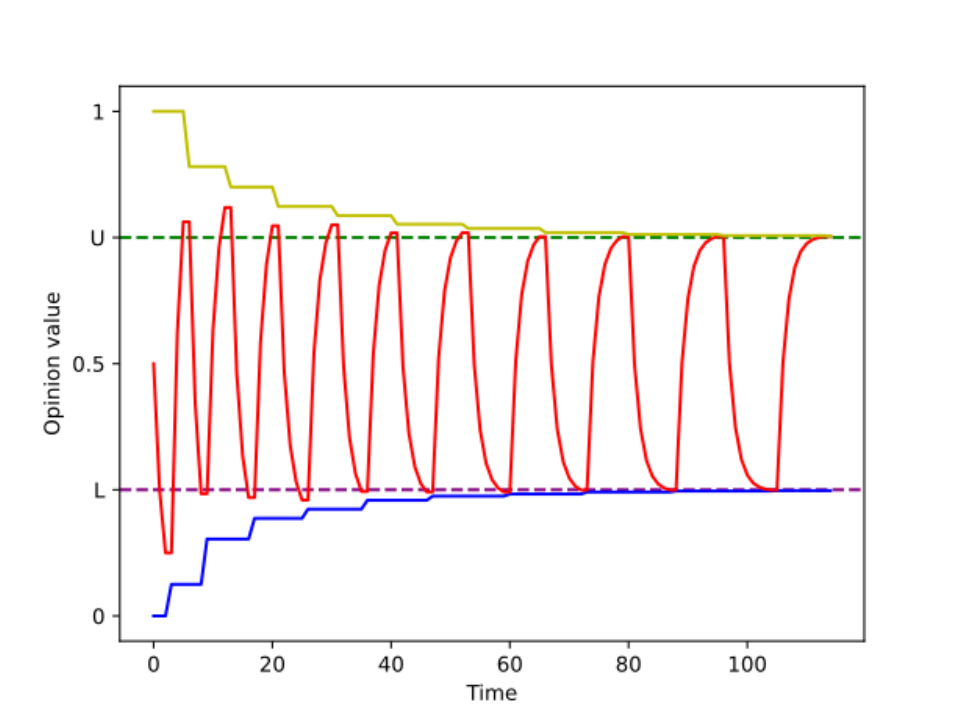}
    \caption{\label{fig:edgeFairnessCounterex2} Opinion evolution of the OTS from Fig. \ref{fig:ex1} for $U=0.75$, $L=0.25$ and the $\omega$-word $w$ from  Cons. \ref{ex:edgeFairnessCounterEx}.}.
    \end{subfigure}

    \caption{Run examples for OTS in Fig. \ref{fig:ex1}.}
    \label{fig:edgeFairnessCounterex}
    
    \end{figure}

Notice that the graph from Ex. \ref{ex:sec:model} is strongly connected and puppet free, and the $\omega$-word $w=(abcd)^\omega$ is indeed strongly fair and converges to consensus. Nevertheless, puppet freedom, strong fairness, and strong connectivity are not sufficient to guarantee consensus.

\begin{proposition}\label{prop:counter-example} There exists $(\graph,\binit,\to)$, where  
$\graph$ is 
strongly connected and puppet free, with a strongly-fair run that does not converge to consensus. 
\end{proposition}

The proof of the existence statement in Prop. \ref{prop:counter-example} is given next. 

\begin{construction}[Counter-Example to Consensus]\label{ex:edgeFairnessCounterEx} 
Let $M=(\graph,\binit,\to)$ be an OTS where $\graph$ is the strongly-connected puppet-free influence graph in Fig. \ref{fig:ex1} and $\binit$ is any state of opinion such that $\binit[1]<\binit[2]<\binit[3]$. We have $A=\{1,2,3\}$ and $E=\{a,b,c,d\}$. We construct an $\omega$-word $w$ such that $\pi_w$ does not converge to consensus with the following infinite iterative process.  Let $U$ and $L$ be such that $\binit[1]<L<\binit[2]<U<\binit[3]$.

\emph{Process:} (1) Perform a non-empty sequence of $a$ actions with as many $a$'s as needed until the opinion of Agent 2 becomes smaller than $L$.  (2) Perform the action $b$. (3)  Perform a non-empty sequence of $c$'s  with as many $c$'s as needed until the opinion of Agent 2 becomes greater than $U$. (4)  Perform the action $d$. The result of this iteration is a sequence of the form $a^+bc^+d$. Repeat steps 1–4 indefinitely.

The above process produces the $\omega$-sequence  $w=w_1\cdot w_2\cdot\ldots$ of the form $(a^+bc^+d)^\omega$, where each $w_i=a^{n_i}bc^{m_i}d$ is the result of the $i$-th iteration of the process and $n_i>0$ and $m_i>0$ are the number of $a$'s and $c$'s in such interaction. (The evolution of the opinion of run $\pi_w$, with $U=0.75$, $L=0.25$ and $\binit=(0,0.5,1)$ is illustrated in Fig. \ref{fig:edgeFairnessCounterex2})

Since each action $e\in E$ appears infinitely often in $w$, $w$ is \emph{strongly fair}. Furthermore,  right after each execution of Step 2, the opinion of Agent 1 gets closer to $L$, but it is still smaller than $L$ since the opinion of Agent 2 at that point is smaller than $L$. For symmetric reasons, the opinion of Agent 3  gets closer to $U$, but it is still greater than $U$.
Consequently, the opinion of Agent 1 is always below $L$, while the opinion of Agent 3 is always above $U$ with $L<U$. Therefore, they cannot converge to the same opinion. 
\end{construction}

Another $\omega$-word for the OTS in Fig. \ref{fig:ex1} exhibiting a behavior similar to $w$ in Cons. \ref{ex:edgeFairnessCounterEx}, but whose proof of non-convergence to consensus seems more involved, is $u= (a^nbc^nd)_{n\in\Nat^+}$= $u_1\cdot u_2\cdot\ldots$, where each $u_n=a^nbc^nd.$  (see Fig. \ref{fig:edgeFairnessCounterex1}).   
The \emph{delay} in both $w$ and $u$ to execute $d$ after $b$ grows unboundedly due to the growing number of $c$'s. More precisely, let $\#e(v)$ be the number of occurrences of $e\in E$ in a finite sequence $v$.

\begin{restatable}{proposition}{unboundedGrowth}\label{prop:unbounded-growth}
Let $w=w_1\cdot w_2\cdot\ldots$ be the $\omega$-word from Cons. \ref{ex:edgeFairnessCounterEx} where each $w_m$ has the form $a^+bc^+d$. Then for every $m \in\Nat$, there exists $t \in \Nat$ such that $\#c(w_{m+t})>\#c(w_{m}).$
    
\end{restatable}

The above proposition states that the number of consecutive $c$'s in $w$ grows \emph{unboundedly}, and hence so does \emph{the delay for executing $d$ right after executing $b$}. To prevent this form of \emph{unbounded delay}, we recall in the next section some notions of fairness from the literature that require, at each position of an $\omega$-word, every action to occur within some bounded period of time.

\subsection{Bounded Fairness}

We start by introducing some notation to give a uniform presentation of some notions of fairness from the literature. We assume $|E|>1$; otherwise, all the fairness notions are trivial.

A word $w$ is a possibly infinite sequence over $E$. 
A \emph{subword} of $w$ is either a suffix of $w$ or a prefix of some suffix of $w$. Let $\kappa$ be an ordinal from the set $\omega+1 = \Nat \cup \{ \omega \} $  where $\omega$ denotes the first infinite ordinal. A \emph{$\kappa$-word} is a word of length $\kappa$.  Recall that each ordinal can be represented as the set of all strictly smaller ordinals. We can then view a $\kappa$-word $w=(e_i)_{i \in \kappa}$ as a function $w:\kappa \to E$  such that $w(i)=e_i$ for each $i \in \kappa$.  A $\kappa$-word $w$ is \emph{complete} if $w(\kappa)=E$ (where $w(\kappa)$ denotes the image of the function $w$). A \emph{$\kappa$-window} $u$ of $w$ is a \emph{subword} of $w$ of length $\kappa$. Thus, if $\kappa=\omega$ then $u$ is a suffix of $w$, and if $\kappa \in \Nat$, $u$ can be thought of as a \emph{finite observation} of $\kappa$ consecutive edges in $w$. We can now introduce a general notion of fairness parametric in $\kappa$. 

\begin{definition}[$\kappa$-fairness, bounded-fairness] 
Let $w$ be an $\omega$-word over $E$ and $\kappa\in \omega+1$: $w$ is \emph{$\kappa$-fair} if every $\kappa$-window of $w$ is complete. Furthermore, $w$ is \emph{bounded fair} if it is $k$-fair for some $k \in \Nat$. 
\end{definition}

Notice that the notion of strong fairness in Def. \ref{def:strongfairness} is obtained by taking $\kappa=\omega$; indeed, $w$ is $\omega$-fair iff  every $e\in E$ occurs infinitely often in $w$. Furthermore, if $\kappa = k$ for some $k\in\Nat^+$, then we obtain the notion of $k$-fairness from \cite{Dershowitz:03}\footnote{This notion is different from the notion of $k$-fairness from \cite{Best1984}}. 
Intuitively, if $w$ is $k$-fair, then at any position of $w$, every $e\in E$ will occur within a window of length $k$ from that position. 


It is not difficult to see that $\omega$-fairness is strictly weaker than bounded-fairness, which in turn is strictly weaker than any $k$-fairness with $k\in\Nat$. 
Let $F(\kappa)$ be the set of all $\omega$-words over $E$ that are $\kappa$-fair. We have the following sequence of strict inclusions. 
\begin{restatable}{proposition}{fairnesshierarchy}\label{prop:k-fairness-hierarchy} For every $k\in\Nat$, 
$F(k)\subset F(k+1) \subset (\bigcup_{\kappa\in\Nat}F(\kappa))\subset F(\omega).$
\end{restatable}

\begin{example} 
Let us consider the \emph{fair word} $w$ from Cons. \ref{ex:edgeFairnessCounterEx}, the  counter-example to consensus. From Prop. \ref{prop:unbounded-growth}, the delay for executing action $d$ immediately after executing action $b$  increases without bound.
Thus, for every $k$, there must be a non-complete $k$-window $u$ of $w$ such that $d$ does not occur in $u$. Consequently, $w$ is not bounded fair. 
\end{example}

Not only does bounded fairness rule out the counter-example in Cons. \ref{ex:edgeFairnessCounterEx}, but it also guarantees consensus, as shown later,  for runs of OTS with strongly-connected, puppet-free influence graphs. Nevertheless, it may be too strong of a requirement for consensus. We, therefore, introduce a weaker notion that satisfies the following criteria and guarantees consensus.

\subsubsection*{Some Fairness Criteria}
Let us briefly discuss some fairness criteria and
desirable properties that justify our quest for a weaker notion of fairness that guarantees consensus. An in-depth discussion about criteria for fairness notions, from which we drew some inspiration, can be found in \cite{Varacca2005concur, Varacca2012,  Glabbeek, Apt1987feasibility}.

{\it\bf Machine Closure}. Following \cite{Lamport1992, Lamport2000} one of the most important criteria that a notion of fairness must meet is \emph{machine closure} (also called \emph{feasibility} \cite{Apt1987feasibility}). Fairness properties are properties of infinite runs; hence, a natural requirement is that any finite partial run must have the chance to be extended to a fair run. Thus, we say that a notion of fairness is \emph{machine closed} if every finite word $u$ can be extended to a fair $\omega$-word $u \cdot w.$ 

Clearly, $k$-fairness with $k\in \Nat$ is not machine closed; e.g., the word $c^kd$ with $E=\{c,d\}$ cannot be extended to a $k$-fair $\omega$-word. Nevertheless, bounded fairness is machine closed:  Each $k$-word $u$ can be extended to a ($k+m$)-fair word $u\cdot(e_1\ldots e_m)^\omega$ assuming $E=\{e_1,\ldots,e_m\}.$

{\it\bf Constructive Liveness.} According to \cite{Varacca2005concur}, a notion of fairness may require that a particular action is taken sufficiently often, but it should not prevent any other actions from being taken sufficiently often. This concept is formalized in \cite{Varacca2005concur, Varacca2012} in a game-theoretical scenario, reminiscent of a Banach-Mazur game \cite{BanachMazur}, involving an infinite interaction between a scheduler and
an opponent. The opponent initiates with a word $w_0$, then the scheduler appends a finite word $w_1$ to $w_0$. This pattern continues indefinitely, resulting in an $\omega$-word $w=w_0\cdot w_1 \cdot w_2\ldots$.  A given fairness notion is said to be a \emph{constructive liveness} property if, regardless of what the opponent does, the scheduler can guarantee that the resulting $\omega$-word is fair under the given notion.

The notion of \emph{bounded fairness is not a constructive liveness property}. If an $\omega$-word is bounded fair, it is $k$-fair for some $k\geq|E|>1$. Let $c \in E$ and take as the strategy of the opponent to choose in each of their turns $w_n=c^n$. Since $|E|>1$, then $w_{2k}$ cannot be a complete $k$-window. Therefore, the resulting $w=w_0\cdot w_1 \cdot w_2\ldots$ is not bounded fair, regardless of the strategy of the scheduler. 

It is worth noticing that the above opponent's strategy is reminiscent of our procedure to construct an $\omega$-sequence in Cons. \ref{ex:edgeFairnessCounterEx} using the unbounded growth of $c$'s to prevent consensus.

{\it\bf Random Words.}  Consider a word $e_0e_1\ldots$ where each edge or action $e_n=(i,j)$ is chosen from $E$ \emph{independently} with some non-zero probability $p_{(i,j)}$. Let us refer to such kinds of sequences as \emph{random} words. We then say that a given notion of fairness is \emph{random inclusive} if every random $\omega$-word is \emph{almost surely} (i.e., with probability one) fair under the given notion.  

It follows from the Second Borel–Cantelli lemma\footnote{The lemma states that if the sum of the probabilities of an infinite sequence of events that are independent is infinite, then the probability of infinitely many of those events occurring is 1. \cite{BorelCantelli}} that every random word is \emph{almost surely} strongly fair.  Nevertheless, the notion of bounded fairness fails to be random inclusive: If a word is bounded fair, it is $k$-fair for some $k\geq|E|$, and thus it needs to have the form $w_0\cdot  w_1 \dots$ where each $w_m$ is a complete $k$-window. Since $1< |E|$, the probability that a random $k$ window is complete is strictly smaller than 1. Therefore, the probability of a random word having an infinite number of \emph{consecutive} complete $k$-windows is 0. 

Random words are important in simulations of our model (see Fig. \ref{fig:introEx3}). Furthermore, having a notion of fairness that is random inclusive and guarantees consensus will allow us to derive and generalize consensus results for randomized opinion models such as gossip algorithms \cite{Fagnani2008}. We elaborate on this in the related work. We now introduce our new notion of fairness.

\section{A New Notion of Bounded Fairness}
\label{section:m-fairness}
A natural way to relax bounded fairness to satisfy constructive liveness and random inclusion is to require that the complete $k$-windows need only appear infinitely often: i.e., an $\omega$ word $w$ is said to be \emph{weakly bounded} fair if there exists $k\in\Nat$ such that every suffix of $w$ has a $k$-window. Nevertheless, as it will be derived later, weak bounded fairness is not sufficient to guarantee consensus. 

It turns out that, to guarantee consensus, it suffices to require that a large enough number $m$ of \emph{consecutive} complete $k$-windows appear infinitely often. These consecutive windows are referred to as multi-windows.

\begin{definition}[$(m,\kappa)$ multi-window]\label{def:multi-window} Let $w$ be an $\omega$-word over $E$, $m\in\Nat^+$ and $\kappa \in \omega+1$. We say that $w$ \emph{has an $(m,\kappa)$ multi-window} if there exists a subword $u$ of $w$ of the form $u=w_1\cdot w_2 \cdot \ldots \cdot w_m$ where each $w_i$ is a $\kappa$-window of $w$. 
Furthermore, if each $w_n$ in $u$ is complete, we say that $w$ \emph{has a complete} $(m,\kappa)$ multi-window. If it exists, the word $u$ is called an \emph{$(m,\kappa)$ multi-window of $w$}. 
\end{definition}

Notice that, because of the concatenation of windows in Def. \ref{def:multi-window}, by construction, no $\omega$-word has a $(m,\omega)$ multi-window with $m>1$: If $\kappa=\omega$ then $m=1$. In this case,
the multi-window is just a window of infinite length of $w$, i.e., a suffix of $w$.

\begin{definition}[$(m,\kappa)$-fairness] Let $w$ be an $\omega$-word over $E$, $m\in\Nat^+$ and $\kappa \in \omega+1$. We say that $w$ is \emph{$(m,\kappa)$-fair} if every suffix of $w$ has a complete $(m,\kappa)$ multi-window. 
We say that $w$ is \emph{$m$-consecutive bounded fair}, or  \emph{$m$-bounded fair},  if it is $(m,k)$-fair for some $k\in\Nat.$
\end{definition}

Clearly, $w$ is $\omega$-fair iff it is $(1,\omega)$-fair, and $w$ is weakly bounded fair iff it is 1-bounded $\omega$-fair. Let $F(m,\kappa)$ and $F(\kappa)$ be the sets of $\omega$-words that are $(m,\kappa)$-fair and $\kappa$-fair, respectively. We have the following sequence of strict inclusions (assume  $k,m\in\Nat^+$):

\begin{restatable}{proposition}{mfairnesshierarchy}\label{prop:m-fairness-hierarchy}  $F(k)\subset F(m+1,k)\subset F(m,k)\subset (\bigcup_{\kappa \in \Nat} F(m,\kappa))\subset F(1,\omega)=F(\omega).$ 
\end{restatable}

{\bf Compliance with Fairness Criteria}. Let us consider the criteria for fairness in the previous section. The  notion of $m$-bounded fairness is machine closed since bounded fairness is stronger than $m$-bounded fairness (Prop. \ref{prop:k-fairness-hierarchy} and Prop. \ref{prop:m-fairness-hierarchy}) and bounded fairness is machine closed. 

 It is also a constructive liveness property since $(m,k)$ fairness, for $k\geq|E|$, is stronger than $m$-bounded fairness (Prop. \ref{prop:m-fairness-hierarchy}), and it is also a constructive liveness property: A winning strategy for the scheduler is to choose a complete $(m,k)$-window at each one of its turns.
 
 Similarly, $m$-Bounded Fairness is random inclusive since the stronger notion $(m,k)$-Fairness is random inclusive for $k\geq|E|$.
 In a random $\omega$-word $w=w_0 \cdot w_1 \dots $ where each $w_n$ is a $(m\times k)$-window, the probability that $w_n$ is a complete $(m,k)$-multi-window is  non-zero and independent. Thus again, by the Second Borel–Cantelli lemma, almost-surely $w$ has infinitely many complete $(m,k)$ multi-windows, i.e., it is almost-surely $(m,k)$-fair.

\subsection{Consensus Theorem}

We can now state one of our main theorems: $m$-bounded fairness guarantees consensus in strongly-connected, puppet-free graphs.

\begin{theorem}[Consensus under $m$-bounded  fairness]\label{th:consensus-bounded-m-fairness}
Let $M=(\graph,\binit,\to)$ be an OTS where $G$ is a strongly-connected, puppet-free influence graph. For every run $\pi$ of $M$, if $w_\pi$ is $m$-bounded fair and $m\geq|A|-1$, then $\pi$ converges to consensus. 
\end{theorem}

\begin{remark} A noteworthy corollary of Th. \ref{th:consensus-bounded-m-fairness} is that, under the same assumptions of the theorem,  if $w_\pi$ is a bounded fair (a random $\omega$-word), then $\pi$ converges to consensus ($\pi$ almost surely converges to consensus).  This follows from the above theorem, Prop. \ref{prop:k-fairness-hierarchy}, Prop. \ref{prop:m-fairness-hierarchy} and the fact that $m$-bounded fairness is random inclusive.
\end{remark}
A proof of Th. \ref{th:consensus-bounded-m-fairness} is given in Section \ref{sect:technicalDetails} in the Appendix. Let us give the main intuitions here.   The proof focuses on the evolution of maximum and minimum opinion values. The sequences of maximum and minimum opinion values in a run, $\{\max \B{t}{}\}_{t\in{\Nat}} $ and $\{\min \B{t}{}\}_{t\in{\Nat}} $, can be shown to be (bounded) monotonically non-increasing and non-decreasing, respectively, so they must converge to some opinion values, say $U$ and $L$ with $L\leq U$. 
 
We must then argue that $L=U$ (this implies convergence to consensus of $\pi$ by the Squeeze Theorem \cite{RealAnalysis}). Since $w_\pi$ is $m$-bounded fair with $m\geq|A|-1$, after performing all the actions of an $(m,k)$ multi-window of $w_\pi$, for some $k\geq |E|$, all the agents of $A$ would have influenced each other. In particular, the agents holding the maximum and minimum opinion values, say agents $i$ and $j$. To see this, notice that since $G$ is strongly connected, there is a path from $i$ to $j$, $a_1 \ldots a_l$ with length $l\leq |A|-1$. Thus, after performing the first complete $k$-window of the $(m,k)$-multi-window, $a_1$ must be performed, after performing the second complete $k$-window,   $a_2$ must be performed and so on. Hence, after performing all the actions of the multi-window, $i$ would have influenced $j$. It can be  shown that their mutual influence causes them to decrease their distance by a positive constant factor (here, the puppet freedom assumption is needed). Since the $w_\pi$ is $m$-fair, there are infinitely many $(m,k)$-windows to be performed, and thus the sequences of maximum and minimum opinion values converge to each other, i.e., $U=L$. \qed

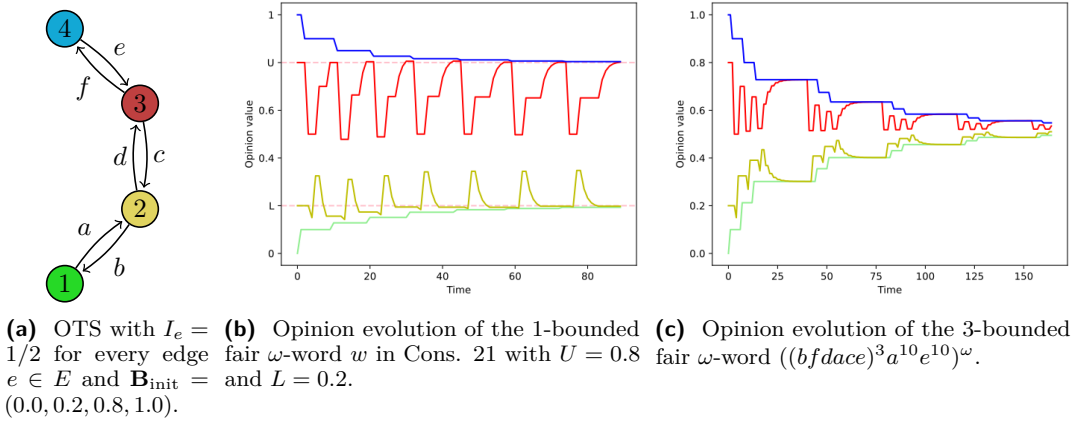
\begin{figure}[t]
\centering%
\begin{subfigure}[t]{0.19\textwidth}%
\centering%
    \begin{tikzpicture}[->, shorten >=1pt,node distance=1.4cm,semithick,
    inner sep=2pt,bend angle=45, initial text=,]
      \node[state, fill=green!70!gray] (1) {1};
      \node[state, fill=yellow!70!gray] (2) [above right of=1] {2};
    \node[state, fill=red!50!gray] (3) [above of=2] {3};
      \node[state, fill=cyan!70!gray] (4) [above left of=3] {4};
    
    \path(1) edge[right, anchor=south east,  bend left=10] node{$a$}
    (2);

    \path(2) edge[left, anchor=north west,  bend left=10] node{$b$}
    (1);

    \path(2) edge[right, anchor=east,  bend left=10] node{$d$}
    (3);

    \path(3) edge[left, anchor=west,  bend left=10] node{$c$}
    (2);

    \path(3) edge[right, anchor=north east,  bend left=10] node{$f$}
    (4);
    
    \path(4) edge[left, anchor=south west,  bend left=10] node{$e$}
    (3);

    \end{tikzpicture}

\caption{\label{fig:mCBFCounterexNetwork} OTS with $I_e = 1/2$ for every edge $e\in E$ and $\binit=(0.0,0.2,0.8,1.0)$.}
\end{subfigure}%
\hfill
\begin{subfigure}[t]{0.39\textwidth}%
\centering%
 \includegraphics[width=1.1\textwidth]{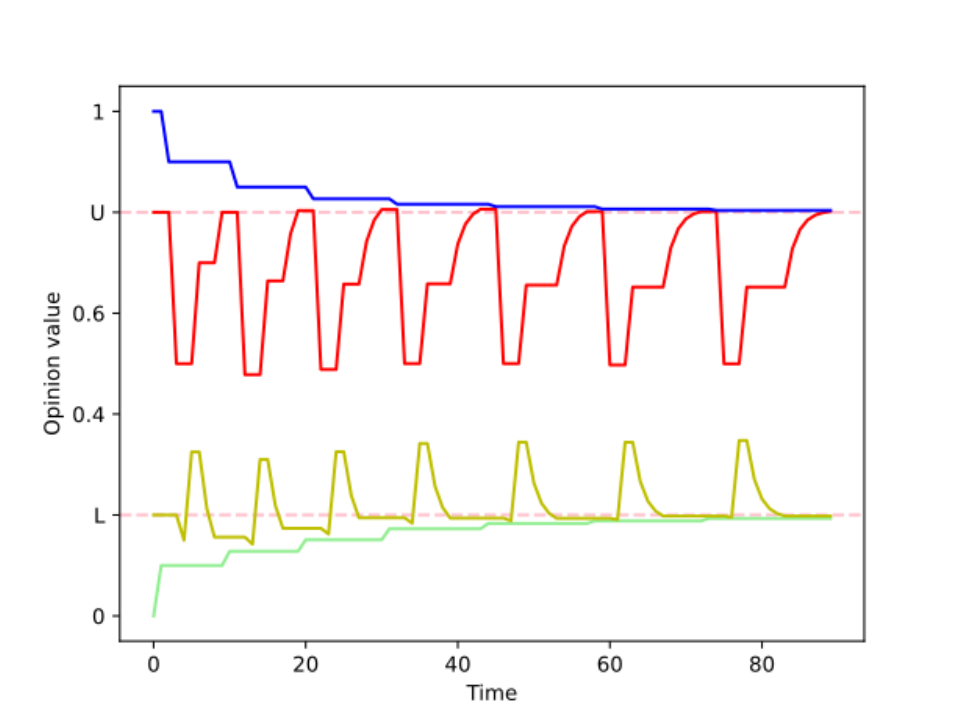}
\caption{\label{fig:mCBFCounterexPlot} Opinion evolution of the $1$-bounded fair $\omega$-word $w$ in Cons. \ref{ex:mbfCounterEx} with $U=0.8$ and $L=0.2$.}
\end{subfigure}%
\hfill
\begin{subfigure}[t]{0.39\textwidth}%
\centering%
 \includegraphics[width=1.1\textwidth]{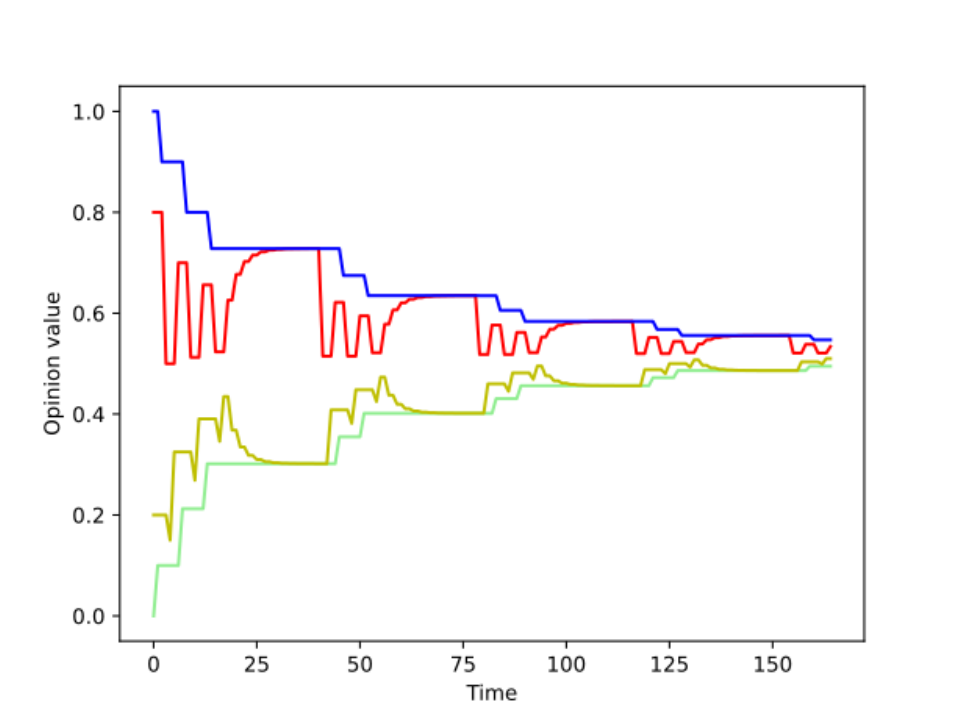}
\caption{\label{fig:mCBFexPlot} Opinion evolution of the $3$-bounded fair $\omega$-word $((bfdace)^3 a^{10}e^{10})^\omega$.}
\end{subfigure}

\caption{\label{fig:mCBFCounterex} Examples of an $m$-bounded fair runs. In Fig. \ref{fig:mCBFCounterexPlot} and \ref{fig:mCBFexPlot}, each plot corresponds to the opinion of the agent with the same color in Fig. \ref{fig:mCBFCounterexNetwork}.
}

\end{figure}

It is worth pointing out that without the condition $m \geq |A|-1$ in  Th. \ref{th:consensus-bounded-m-fairness}, we cannot guarantee consensus. Fig. \ref{fig:mCBFexPlot} illustrates an $m$-bounded fair run, for $m=|A|-1$, of an OTS with 4 agents that converges to consensus. Nevertheless, the following run construction shows that for $m=|A|-3$, we can construct an $m$-bounded fair run that fails to converge to consensus (the run is illustrated in Fig. \ref{fig:mCBFCounterexPlot}). It also shows that weak bounded fairness, i.e., $1$-bounded fairness, is not sufficient to guarantee convergence to consensus. We do not have a counter-example or a proof for $m=|A|-2$.

\begin{proposition}\label{prop:mBF-counter-example} There exists $M=(\graph,\binit,\to)$, where  
$\graph=(\agents,\edges,\I{})$ is a
strongly connected, puppet-free graph, with an $m$-bounded fair $\omega$-word $w$, $m=|A|-3$, such  that $\pi_w$ does not converge to consensus. 
\end{proposition}

The proof of the above proposition is given in the following construction.

\begin{construction}[Counter-Example to Consensus for $m$-bounded fairness with $m \leq |\agents| - 3$]\label{ex:mbfCounterEx} Suppose that $M=(\graph,\binit,\to)$ where $\graph$ is the strongly-connected, puppet-free, influence graph in Fig. \ref{fig:mCBFCounterexNetwork} and $\binit$ is any state of opinion such that $\binit[1]<\binit[2]<\binit[3]<\binit[4]$. We have $A=\{1,2,3,4\}$ and $E=\{a,b,c,d,e,f\}$. We construct an $\omega$-word $w$ such that $\pi_w$ does not converge to consensus with the following infinite iterative process. Let $U$ and $L$ be such that $\binit[2]\leq L<U\leq\binit[3]$.

\emph{Process:} 
(1) Perform the sequence of actions $\mathit{bfdace}$. (2) Perform a sequence of $a$ actions with as many $a$'s as needed until the opinion of Agent 2 becomes smaller than $L$. (3)  Perform a sequence of $e$'s  with as many $e$'s as needed until the opinion of Agent 3 becomes greater than $U$. The result of this iteration is a sequence of the form $\mathit{bfdace}\cdot a^*e^*$. Repeat steps 1-3 indefinitely.

The above process produces the $\omega$-sequence $w=v\cdot w_1\cdot v\cdot w_2\cdot\ldots$ of the form $(\mathit{bfdace} \, a^*e^*)^\omega$ where $v=\mathit{bfdace}$ and $w_i=a^{m_i}e^{n_i}$ are results of the $i$-th iteration of the process, and $n_i\geq 0$ and $m_i \geq 0$ are the number of $a$'s and $e$'s in each $w_i$. (The opinion evolution of run $\pi_w$, with $L=0.2$, $U=0.8$ and $\binit=(0.0,0.2,0.8,1.0)$ is illustrated Fig. \ref{fig:mCBFCounterexPlot})

Since the subword $v$ is a complete (1,6)-multi-window and appears infinitely often in $w$, $w$ is \emph{$m$-bounded fair} for $m=|\agents|-3=1$. Furthermore, right after each execution of edge $f$ in step 1, the opinion of Agent 1 gets closer to $L$, but it is  still smaller than $L$ since the opinion of Agent 2 at that point is smaller than $L$. For symmetric reasons, after action $b$, the opinion of Agent 4 gets closer to $U$, but it is still greater than $U$ since the opinion of Agent 3 at that point is greater than $U$.
Consequently, the opinion of Agent 1 is always below $L$, while the opinion of Agent 4 is always above $U$ with $L<U$. Therefore, they cannot converge to the same opinion. 

\end{construction}%

\section{Dynamic Influence}
\label{sect:dynamicInfluence}

\begin{figure}[t]
\centering%
\begin{subfigure}[t]{0.45\textwidth}%
\centering%
    \begin{tikzpicture}[->, shorten >=1pt,node distance=2cm,semithick,
    inner sep=2pt,bend angle=45, initial text=,]
      \node[state, fill=cyan!70!gray] (1) {2};
      \node[state, fill=red!50!gray] (2) [right of=1] {1};

    \path(1) edge[right, anchor=south ,  bend left=10] node{$a$}
    (2);

    \path(2) edge[left, anchor=north ,  bend left=10] node{$b$}
    (1);
    
    \end{tikzpicture}

\caption{\label{fig:dynamicCounterexNetwork1} $\binit=(0.0,1.0)$ and if $\B{}{1}=\B{}{2}$ then $\IB{a} =\IB{b}=0.5$, otherwise 
\\
$\IB{a} = \left[ \frac{U- \B{}{2}}{2 (\B{}{1}-\B{}{2})} \right]_0^1, \IB{b} = \left[ \frac{L- \B{}{1}}{2 (\B{}{2}-\B{}{1})}\right]_0^1$
}
\end{subfigure}%
\hfill%
\begin{subfigure}[t]{0.54\textwidth}%
\centering%
    \begin{tikzpicture}[->, 
    shorten >=1pt,
    node distance=2cm,
    semithick,
    inner sep=2pt,
    bend angle=45, 
    initial text=,]
      \node[state, fill=yellow!70!gray] (1) {3};
      \node[state, fill=red!50!gray] (2) [right of=1] {2};
    \node[state, fill=cyan!70!gray] (3) [right of=2] {1};
    
    \path(1) edge[right, anchor=south,  bend left=10] node{$c$}
    (2);

    \path(2) edge[left, anchor=north,  bend left=10] node{$d$}
    (1);

    \path(2) edge[right, anchor=south,  bend left=10] node{$b$}
    (3);

    \path(3) edge[left, anchor=north,  bend left=10] node{$a$}
    (2);

    \end{tikzpicture}

\caption{\label{fig:dynamicCounterexNetwork2} $\binit=(0.0,0.5,1.0)$, $\IB{d} = \IB{b} = 0.5$, if $\B{}{1}=\B{}{2}$ then  $\IB{a}=0.5$, if $\B{}{2}=\B{}{3}$ then $\IB{c}=0.5$, otherwise \\ $\IB{a} = \left[ \frac{\frac{1}{2}(\B{}{1}+L)-\B{}{2}}{\B{}{1}-\B{}{2}}\right]_0^1 \text{, } \IB{c} = \left[ \frac{\frac{1}{2}(\B{}{3}+U)-\B{}{2}}{\B{}{3}-\B{}{2}}\right]_0^1$
}
\end{subfigure}

\begin{subfigure}[t]{0.49\textwidth}%
\centering%
\includegraphics[width=0.8\textwidth]{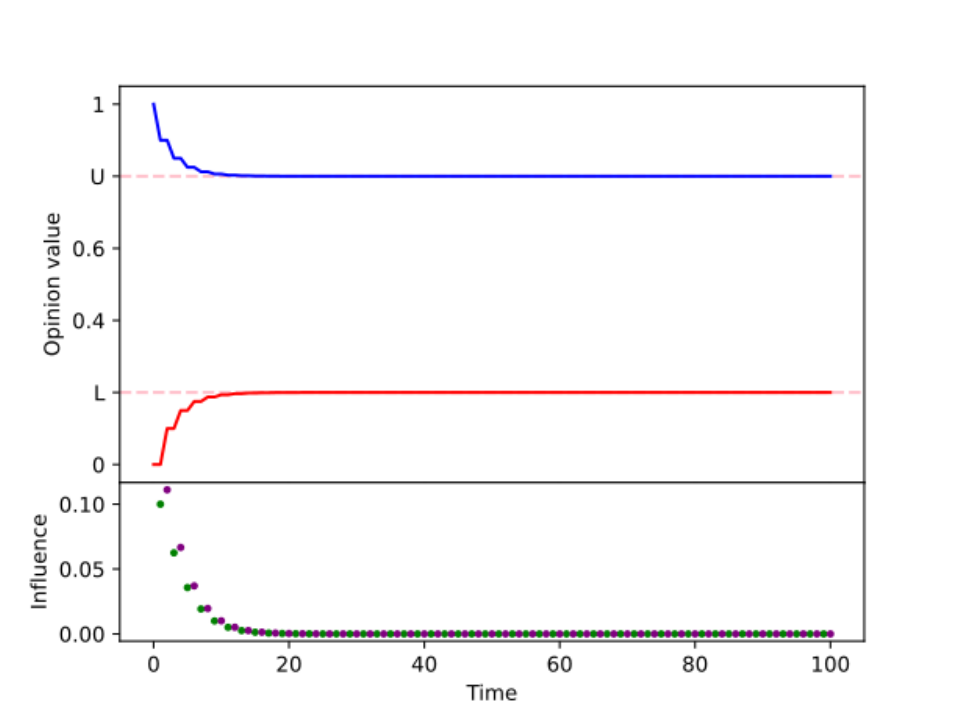}
\caption{\label{fig:dynamicCounterexPlot1} Opinion and influence evolution of the $\omega$-word $(ab)^\omega$. Each plot corresponds to the opinion of the agent with the same color in Fig. \ref{fig:dynamicCounterexNetwork1}. The influences $\IB{a}$ and $\IB{b}$ are plotted in green and purple.}
\end{subfigure}%
\hfill%
\begin{subfigure}[t]{0.49\textwidth}%
\centering%
\includegraphics[width=0.8\textwidth]{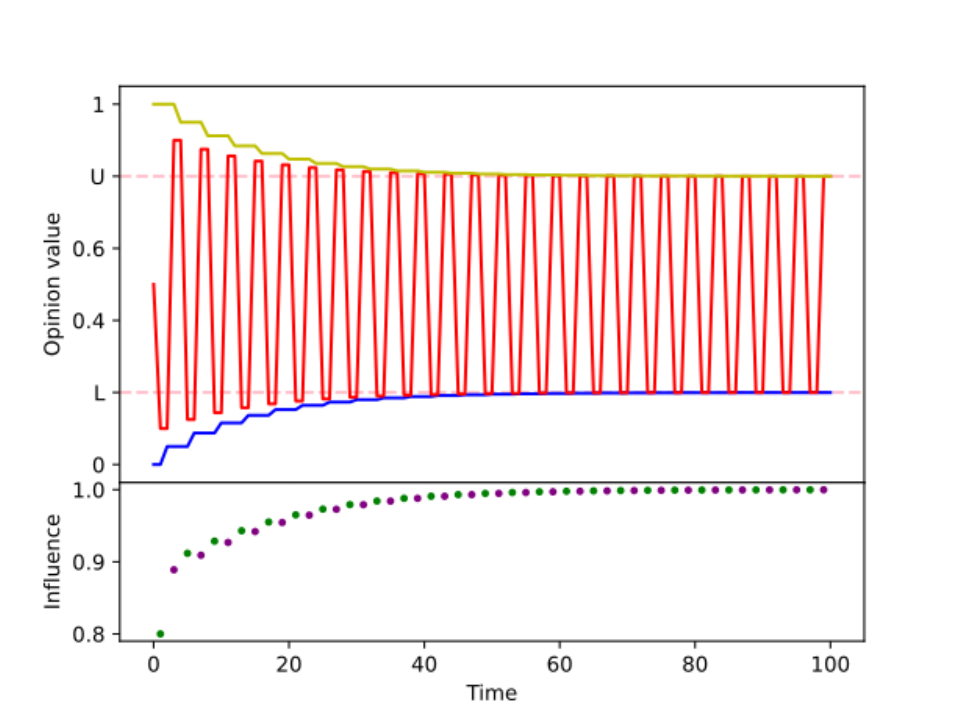}
\caption{\label{fig:dynamicCounterexPlot2} Opinion and influence evolution of the $\omega$-word $(a\, b\, c\, d)^\omega$. Each plot corresponds to the opinion of the agent with the same color in Fig. \ref{fig:dynamicCounterexNetwork2}. The influences $ \IB{a}$ and $\IB{c}$ are plotted in green and purple.}
\end{subfigure}

\caption{\label{fig:dynamicCounterex}Plots for DOTS in Fig. \ref{fig:dynamicCounterexNetwork1} and Fig. \ref{fig:dynamicCounterexNetwork2}
with $U=0.8$ and $L=0.2$.\protect\footnotemark}
\end{figure}

\footnotetext{We use a clamp function for $[0,1]$ defined as $[r]_0^1 = \min(\max(r,0),1)$ for every $r\in\mathbb{R}$.}

 The static weights of the influence graph of an OTS imply that the influence that each individual has on others remains constant throughout opinion evolution. However, in real-life scenarios, the influence of individuals can vary depending on many factors, in particular the state of opinion (or opinion climate). Indeed, individuals may gain or lose influence based on the current opinion trend or for expressing dissenting and extreme opinions, among others.

 To account for the above form of dynamic influence, we extend the weight function $I:E\to (0,1]$ of the influence graph $\graphElem$ as a function $I:E\times [0,1]^{|A|} \to [0,1]$ on edges and the state of opinion. The resulting graph is said to have \emph{dynamic influence}.

\begin{definition}[Dynamic OTS]\label{def:dots}
    A Dynamic OTS (DOTS) is a tuple $\model$ where $\graphElem$ has dynamic influence $\I{}: \edges\times [0,1]^{|A|} \to [0,1]$. We write $\IB{ij}$ for $I((i,j),\B{}{})$. The labeled transition $\to$ is defined as in Def. \ref{def:ots} but replacing $\I{ij}$ with $\IB{ij}$ in Eq. \ref{eq:reduction}.
\end{definition}

The notions of runs, words, e-paths, and related notions for DOTS remain the same as those for OTS (Def. \ref{def:runs}). Let us consider some examples of dynamic influence. 

{\bf Confirmation Bias}. Under \emph{confirmation bias} \cite{Aronson2010sociology}, an agent $j$ is more influenced by those whose opinion is closer to theirs.  The function $\IB{ij} = 1 - \lvert \ \B{}{j}-\B{}{i}\ \rvert $ captures a form of confirmation bias; the closer the opinions of $i$ and $j$, the stronger the influence of $i$ over $j$.

{\bf Bounded Influence.} Nevertheless, if we allow dynamic influence that can converge to 0 in a given run $\run$, i.e,  if $\lim_{t\to \infty}{\I{i,j}^{\B{t}{}}}=0$, we may reduce indefinitely influence and end up in a situation similar to non-strong connectivity of the graph, thus preventing consensus as in  Section \ref{sec:strong-connectivity} (Fig. \ref{fig:strConnCounterex}). Analogously,  if $\lim_{t\to \infty}{\I{i,j}^{\B{t}{}}}=1$,    
we may end up in puppet situations preventing consensus like in Section \ref{sect:puppetFreedom} 
 (Fig. \ref{fig:puppets}). Both situations are illustrated in the DOTS in 
 Fig. \ref{fig:dynamicCounterex}. 
To prevent them, we bound the dynamic influences.

\begin{definition}[Bounded Influence] A DOTS $\model$ with $\graphElem$\ \emph{has bounded influence} if there are constants $\I{L},\I{U}\in(0,1)$ such that
for each $\B{}{} \in [0,1]^{|A|}$, $(i,j) \in \edges$, we have $\IB{i,j} \in [\I{L},\I{U}]$.
\end{definition}

The previous form of confirmation bias influence $\IB{ij} = 1 - |\B{}{j}-\B{}{i}|$ is not bounded.  Nevertheless, the linear transformation $\I{L} + ( \I{U} - \I{L})\IB{ij}$ can be used to scale any unbounded influence $\IB{ij}$ into a bounded one in $[\I{L},\I{U}]$ while preserving its shape.

We conclude with our other main theorem whose proof is given in the Appendix.




\begin{restatable}[Consensus with bounded influence]{theorem}{dynamicConsensus}

Let $M=(\graph,\binit,\to)$ be a DOTS where $G$ is a strongly-connected, influence graph. Suppose that $M$ has bounded influence. For every run $\pi$ of $M$, if $w_\pi$ is $m$-bounded fair with $m\geq|A|-1$, then $\pi$ converges to consensus. 
\end{restatable}

The result generalizes Th. \ref{th:consensus-bounded-m-fairness} to dynamic bounded influence. Therefore, in strongly-connected and dynamic bounded influence graphs, convergence to consensus is guaranteed for all runs that are $m$-bounded fair, which include each random run almost surely.


\section{Conclusions and Related Work}
\label{sect:RelatedWork}
We introduced a DeGroot-based model with asynchronous opinion updates and dynamic influence using labelled transition systems. The model captures opinion dynamics in social networks more faithfully than the original DeGroot model. The fairness notions studied and the consensus results in this paper show that the model is also tractable and brings new insights into opinion formation in social networks. To our knowledge, this is the first work that uses fairness notions from concurrent systems in the context of social learning.

There is a great deal of work on DeGroot-based models for social learning (e.g.,   \cite{alvim:hal-03740263,Generalize1,Generalize2,Generalize3,GeneralizeBF1,demarzo2003persuasion,chatterjee1977towards}). We discuss work with asynchronous updates and dynamic influence, which is the focus of this paper. The work \cite{demarzo2003persuasion} introduces a version of the DeGroot model in which the self-influence changes over time while the influence on others remains the same. The works \cite{chatterjee1977towards, Generalize2} explore convergence and stability, respectively, in models where influences change over time.  The works mentioned above do not take into account asynchronous communication, whereas this paper demonstrates how asynchronous communication, when combined with dynamic influence, can prevent consensus. 

Recent works on gossip algorithms  \cite{Fagnani2008,Shi2016,Emerico2019,Wang2022} study consensus with asynchronous communications for distributed averaging and opinion dynamics. The work in \cite{Shi2016} studies \emph{reaching} consensus (in finite time) rather than \emph{converging} to consensus. The works \cite{Emerico2019,Wang2022} consider undirected cliques rather than directed graphs as influence graphs. The closest work is  \cite{Fagnani2008}, which states consensus for random runs in directed strongly connected graphs with static influence. The dynamics of asymmetric gossip updates in  \cite{Fagnani2008} can indeed be captured as OTS, and their random runs are almost-surely $m$-bounded fair. Consequently, our work generalizes the consensus result in \cite{Fagnani2008} by extending it with (bounded) dynamic influence.

The work \cite{Glabbeek} discusses probabilistic fairness as a method equally strong as strong fairness to prove liveness properties, where a liveness property is characterized by a set of states such that a run holds this property iff the run reaches a state of this set. However, the property of \emph{(convergence to) consensus}  (Def. \ref{def:consensus})  does not correspond to this notion of liveness since it is not about  \emph{reaching} a specific set of states but about \emph{converging} to a consensual state.  In fact, unless there are puppets or the initial state of a run is already a consensual state, consensus is never reached in finite time in our model.

\bibliography{references}


\appendix
\section{Proof of consensus}\label{sect:technicalDetails}

We prove the consensus of Th. \ref{th:consensus-bounded-m-fairness} in three steps. (1) We show that the maximum and minimum opinion values converge, respectively, to some values $U,\, L \in [0,1]$, $L\leq U$. (2) We establish a conditional proof of $U=L$ by deriving bounds ensuring that the maximum opinion decreases under the influence of agents with the minimum opinion. The conditions are strong connectivity, puppet freedom, and a concept we introduce called \emph{recurrent $\Delta-$bound}. Finally, (3) we show that $m$-bounded fair runs (on  strongly-connected puppet-free graphs) fulfil these conditions for  $m\geq|A|-1$ and therefore ensure consensus.

The complete proofs of the lemmas used next are found in \autoref{sect:proofs}. For simplicity, assume an underlying OTS $M=(\graph,\binit,\to)$ with an strongly-connected puppet-free influence graph $\graph=(\agents,\edges,I)$. 

{\bf Step 1.} We show that the opinion values in a state is bounded by the extreme opinions in the previous state.

\begin{restatable}[Opinion evolution is bounded by the extremes]{lemma}{extremeBounds}\label{lemma:beliefExtremalBounds}
Let $\B{}{} \xrightarrow{e} \B[B']{}{}$ be a transition. Then $min(\B{}{}) \leq \B[B']{}{k} \leq max(\B{}{})$ for all $k \in \agents$.
\end{restatable}


Notice that monotonicity does not necessarily hold for the opinion of agents (e.g., see Fig. \ref{fig:introEx3}). Nevertheless, it follows from lemma \ref{lemma:beliefExtremalBounds} that $max(\B{t}{})$ is monotonically non-increasing and $min(\B{t}{})$ is monotonically non-decreasing with respect to $t$.

\begin{restatable}[Monotonicity of extremes]{corollary}{extremeMonotonicity}\label{corollary:extremeMonotonicity}
     Let $\execution$ be an e-path. $max(\B{t+1}{}) \leq max(\B{t}{})$ and $min(\B{t+1}{}) \geq min(\B{t}{})$  for all $t \in \mathbb{N}$.
 \end{restatable}

Monotonicity and boundedness of extremes, together with the Monotonic Convergence Theorem \cite{RealAnalysis}, lead us to the existence of limits for opinions of extreme agents.

\begin{restatable}[Limits of extremes]{theorem}{extremeLimits}\label{theo:extremeLimits}
    Let $\execution$ be an e-path. There exist $U,\,L \in [0,1]$ such that $\lim_{t \to \infty} \{max(\B{t}{})\} = U$ and $\lim_{t \to \infty} \{min(\B{t}{})\} = L$.
\end{restatable}

Therefore, by the squeeze theorem \cite{RealAnalysis}, to prove Th. \ref{th:consensus-bounded-m-fairness}, it suffices to show that $\{\max(B^t)\}_{t \geq 0}$ and $\{\min(B^t)\}_{t \geq 0}$ converge to the same value. 

{\bf Step 2}. To prove consensus, we now show that $U=L$ for $U,L$ in Th. \ref{theo:extremeLimits}.  We say that an e-path $\pi'$ is an \emph{e-suffix} of an e-path $\pi$ if $\pi'$ is a suffix of $\pi$.
In what follows we let $\pi'=\ \execution$ be an e-suffix of an strongly-fair run and $w=e_0e_1\dots$ be the $\omega$-word generated by $\pi'$.

In Lem. \ref{lemma:opinionSpread}  we identify a lower bound on how much $\{\max(\B{t}{})\}_{t \geq 0}$ decreases. The decrease may occur when an agent of minimum opinion influences all agents.\footnote{A proof of consensus with the maximum opinion influencing all agents can be analogously presented.} To characterize it, we define a function $\Delta_w(i)$ that quantifies how long it takes for an agent to influence every other. This requires some notation.

Recall that  a sequence ($g$-path in this case)  $p=a_0a_1\dots a_n$ is a \textit{subsequence} of a word $w=e_0e_1\dots$ if there exist indices $i_0 < i_1 < \ldots < i_n$ such that $e_{i_0} = a_0$, $e_{i_1} = a_1$, $\ldots$, $e_{i_n} = a_n$. We define $\delta_w(p)$ as the length of the smallest prefix $w'$ of $w$ that such that $p$ is a subsequence of $w'$.

Recall that by definition $g$-paths are finite (i.e., simple paths in the graph). Let $P_\graph(i)$ be the set of all the g-paths starting from $i$.

\begin{definition}[$\Delta$]\label{def:maximalWait}
    For all $i \in \agents$, we define 
     $\Delta_w(i)= \max \{\delta_w(p) \;|\; p \in P_\graph(i)\}$
\end{definition}

Intuitively, $\Delta_w(i)$ is the length of the smallest prefix of $w$ that has all g-paths that start with agent $i$ as subsequences. 
Because $\graph$ is strongly connected, agent $i$ must have influenced every other agent after the $\Delta_w(i)$-th action in $w$.

\begin{example}
    Consider the OTS from Fig. \ref{fig:ex1}. The path $ad$ is a subsequence of $abcd$. It holds that $P_\graph(1)=\{a,\,ad\}$ and $P_\graph(3)=\{c,\,cb\}$. 
    Take the $\omega$-word $w=(abcd)^\omega$. Then, $\delta_w(a)=1$, $\delta_w(ad)=4$, $\delta_w(c)=3$, and $\delta_w(cb)=6$; $\Delta_w(1)=4$ and $\Delta_w(3)=6$. 

    Now take the $\omega$-word $u=(a^nbc^nd)_{n\in\Nat^+}$ from the counter-example to consensus in Fig. \ref{fig:edgeFairnessCounterex1}. 
    Then $\Delta_u(1)=5$ and $\Delta_u(3)=7$. But for the suffix $u'=(a^nbc^nd)_{n\geq10}$ of $u$, $\Delta_{u'}(1)=22$ and $\Delta_{u'}(3)=34$. In this case, $\Delta_{u'}$ increases the later the suffix $u'$ starts in $u$.
\end{example}

The function $\Delta_w(i)$ allows us to express a bound on the maximum opinion in terms of the opinion of $i$ and the constants $max(\B{0}{})$, $|\agents|$ and the maximum and minimum influences of the graph. Let $\I{max}=\max_{(i,j) \in \edges} \I{ij}$ and $\I{min}=\min_{(i,j) \in \edges} \I{ij}$.


\begin{restatable}[Opinion bound over a network]{lemma}{opinionSpread}\label{lemma:opinionSpread}
     Let $\pi'=\execution$ be an e-suffix of a strongly-fair run and $w=e_0e_1\dots$ the $\omega$-word generated by $\pi'$. If $\graph$ is strongly connected, then for all $i \in \agents$
     $$max(\B{0}{}) - max(\B{\Delta_w(i)}{})  \geq \I{min}^{|\agents|}(1-\I{max})^{\Delta_w(i)}(max(\B{0}{})-\B{0}{i})$$
\end{restatable}

This quantifies a decrement in the opinion of the maximum opinion based on the initial opinion of an agent $\B{0}{i}$. However, this decrease may become unboundedly smaller if $\Delta_w(i)$ grows unboundedly, as in Cons. \ref{ex:edgeFairnessCounterEx}. Therefore, we \textit{bound} $\Delta_w(i)$ when $i$ is an agent of minimum opinion. Define $ m_{\pi'} \in \agents$ as the least\footnote{Assume $\agents$ is ordered under the usual order in the natural numbers.} agent in $\agents$ such that $ \B{0}{m_{\pi'}} = min(\B{0}{})$. 

\begin{definition}[$\Delta$-bound]\label{def:boundedWait}
     Let  $\pi$ be a strongly fair run and $\pi'$ an e-suffix of $\pi$. We say $\beta \in \mathbb{N}$ is a $\Delta$-bound of $\pi'$ if $\Delta_{w_{\pi'}}(m_{\pi'}) \leq \beta$.
     
     We say $\beta \in \mathbb{N}$ is a recurrent $\Delta$-bound of $\pi$ if for infinitely many e-suffixes $\pi'$ of $\pi$, $\beta$ is a $\Delta$-bound of $\pi'$.
\end{definition}

Intuitively, a $\Delta$-bound of an e-suffix $\pi'$ is a length bound for the smallest prefix of $w_{\pi'}$ that has all g-paths that start with an agent of minimum opinion as subsequences. A recurrent  $\Delta$-bound of a run $\pi$ is a $\Delta$-bound for infinitely many e-suffixes of $\pi$. With this, we apply \autoref{lemma:opinionSpread} with an agent of minimum opinion to bound the maximum opinion.

\begin{restatable}[$n-\epsilon$ decrement]{lemma}{nepsilon}\label{lemma:nepsilon}
    Let $\pi=\binit\xrightarrow{e_0}\B{1}{}\xrightarrow{e_1}\dots$ be a run with a strongly-connected $\graph$ and a recurrent $\Delta$-bound $\beta \in \mathbb{N}$. For all $n \in \mathbb{N}$, there exists an e-suffix $\pi'=\B{t}{}\xrightarrow{e_t}\B{t+1}{}\xrightarrow{e_{t+1}}\dots$ of $\pi$ such that $\beta$ is a $\Delta$-bound of $\pi'$ where 

    $$max(\binit) - max(\B{t+\beta}{}) \geq   n*\epsilon$$

    with $\epsilon = \I{min}^{|\agents|}(1-\I{max})^{\beta}(U - L)$.
\end{restatable}



   


This is enough to prove $U=L$ by contradiction. Suppose $U \neq L$. We have $\I{max} < 1$ by puppet freedom. Then $\epsilon$ is greater than a positive constant. Using \autoref{lemma:nepsilon}, take $n \in \Nat$ such that $n*\epsilon > max(\binit)$, then $max(\B{t+\beta}{}) < 0$ for some $t$, a contradiction. Therefore, $U=L$.

If $U=L$, the agents of extreme opinion converge to consensus, and by the squeeze theorem \cite{RealAnalysis}, every agent converges to the same opinion. Thus, any run of an OTS with strongly-connected puppet-free $\graph$ and a recurrent $\Delta$-bound converges to consensus.

\begin{restatable}[Consensus with recurrent $\Delta$-bound]{lemma}{deltaBoundConsensus}\label{lemma:deltaBoundConsensus}
    Let $M=(\graph,\binit,\to)$ be an OTS where $G$ is a strongly-connected, puppet-free influence graph. Then for every run $\pi$ of $M$, if there exist $\beta \in \Nat$ such that $\beta$ is a recurrent $\Delta$-bound of $\pi$, then $\pi$ converges to consensus. 
\end{restatable}

{\bf Step 3}. It remains to show that  $m$-bounded fair runs, with $m\geq |A|-1$, have a recurrent $\Delta$-bound.

\begin{restatable}[$m$-bounded fair runs have a recurrent $\Delta$-bound]{lemma}{mbfDeltaBound}\label{lemma:mbfDeltaBound}
    Any bounded $m$-fair run with  $m \geq |\agents| - 1$ has a recurrent $\Delta$-bound $\beta=(|\agents| - 1) \times k$. 
\end{restatable}\todo{JS: Quantify $k$}

The intuition is that if $\pi$ is $m$-bounded fair, it is $(m,k)$-fair for some $k$. This $(m,k)$-fairness provides a complete $(m,k)$ multi-window for every e-suffix. All g-paths must be a subsequence of any complete $(m,k)$ multi-window with $m \geq |\agents| - 1$, because a g-path can visit at most $|A|- 1$ agents. This implies $\Delta_{w_{\pi'}}(m_{\pi'}) \leq \beta = (|\agents| - 1) \times k$ for every e-suffix $w'$ of $w$ that starts with a complete $(m,k)$ multi-window.

Therefore, a bounded $m$-fair run of an OTS with $m \geq |\agents| - 1$, strong connectivity and puppet freedom also has a recurrent $\Delta$-bound, which implies convergence to consensus.

\section{Proofs}\label{sect:proofs}

In this appendix, the reader may find the proofs of the following results:

\begin{itemize}
\item \hyperref[proof:unbounded-growth]{\textsf{\textbf{Proposition 11. }}}

\item \hyperref[proof:fairness-hierarchy]{\textsf{\textbf{Proposition 13.}}}

\item \hyperref[proof:m-fairness-hierarchy]{\textsf{\textbf{Proposition 17.}}}

\item \hyperref[proof:dynamicConsensus]{\textsf{\textbf{Theorem 24} (Consensus with bounded influence).}}

\item \hyperref[proof:extremeBound]{\textsf{\textbf{Lemma 25} (Opinion evolution is bounded by the extremes).}}

\item \hyperref[proof:opinionSpread]{\textsf{\textbf{Lemma 30} (Opinion bound over a network).}}

\item \hyperref[proof:nepsilon]{\textsf{\textbf{Lemma 32} ($n - \epsilon$ decrement).}}

\item \hyperref[proof:deltaBoundConsensus]{\textsf{\textbf{Lemma 33} (Consensus with recurrent $\Delta$-bound).}}

\item \hyperref[proof:mbfDeltaBound]{\textsf{\textbf{Lemma 34} ($m-$bounded fair runs have a recurrent $\Delta$-bound).}}

\item \hyperref[proof:minimumEffort]{\textsf{\textbf{Proposition 35} (Minimum effort for pulling an agent over a given belief).}}

\item \hyperref[proof:greaterStep]{\textsf{\textbf{Lemma 36} (Opinion bound after one step).}}

\item \hyperref[proof:OpinionBoundAfterNSteps]{\textsf{\textbf{Lemma 37} (Opinion upper bound after n steps).}}

\item \hyperref[proof:OpinionBoundOfAdjacentAgent]{\textsf{\textbf{Lemma 38} (Direct influence bound).}}

\item \hyperref[proof:OpinionBoundPath]{\textsf{\textbf{Lemma 39} (Opinion upper bound along a path).}}

\item \hyperref[proof:epsilonDecrement]{\textsf{\textbf{Lemma 41} (epsilon decrement $\Delta$-bound e-suffix).}}

\item \hyperref[proof:epsilonDecrementBoundedWaitRun]{\textsf{\textbf{Lemma 42} (epsilon decrement $\Delta$-bound run).}}
\end{itemize}


\unboundedGrowth*
\begin{proof}
\label{proof:unbounded-growth}

We will  prove this proposition in two steps. 
First (claim 1), for any $m \in Nat$ we will prove, using Prop. \ref{prop:minimumEffort}, that there exists $\mbox{OE} \in \Nat^+$ such that $\#c(w_{m}) \leq \mbox{OE}$. 
Second (claim 2), we will prove, using again Prop. \ref{prop:minimumEffort} and the fact that  $\B{}{3}$ converges to $U$, that there exists $t \in Nat^+$  and $\mbox{UE} \in \Nat^+$  such that $\mbox{OE} < \mbox{UE}$ and $\mbox{UE} \leq \#c(w_{m+t})$.
Then, we can conclude that \emph{for each $m\in \Nat$, there exists $t \in \Nat^+ $ such that  $\#c(w_{m+t}) > \#c(w_{m}) $}.

{\bf Claim 1:}
Let $L, U \in (0,1)$ be the two fixed values used for building $w$, such that $0 < L < \B{0}{2} < U < 1$. 

Let $ua^{\#a(w_{m})}b$ be a prefix of $w$ that represents a partial execution of $w$, that is,  $w=ua^{\#a(w_{m})}bw'$ and $w'=c^{\#c(w_{m})}dw_{m+1}\ldots$

 Let $\B{}{2}$ and $\B{}{3}$ be the opinions of agents 2 and 3 respectively, before starting execution of $w'$. We know that, by definition,  the $\#c(w_{m})$ executions of edge $c$ will carry the opinion of agent 2, just over $U$ and below  $\B{}{3}$. We don't know the exact value of $\#c(w_{m})$, but we can overestimate it easily  estimating the minimum effort needed for carrying a opinion of $0$ over $U$ using the actual opinion of agent 3, $\B{}{3}$. We call this overestimation $\mbox{OE}$. 
 
 Using proposition \ref{prop:minimumEffort} we can conclude that $$\mbox{OE} = \lceil \log_2(\frac{\B{}{3} - 0}{\B{}{3} - U}) \rceil = \lceil \log_2(\frac{\B{}{3}}{\B{}{3} - U}) \rceil$$
 
 Because of $\B{}{2}>0$ we can conclude that $ \mbox{OE} \geq \#c(w_{m})$.

{\bf Claim 2:}
 Now, because of $lim_{t \rightarrow \infty} \B{t}{3} = U$, we  can be sure that executing the $w_{m+1}w_{m+2}\ldots$ run there will be a moment $m+t$, just after executing the $d$ edge of $w_{m+t-1}$, where $\B{}{3} \leq U+\epsilon$ for any $\epsilon \in (0,\B{m+t-1}{3}-U)$. 
 
 That is $w_{1}w_{2} \ldots w_{m}w_{m+1} \ldots w_{m+t-1}$ has been executed and $\B{}{3} \leq U+\epsilon$. Now the edges of $w_{m+t}=a^{p_{m+t}}bc^{q_{m+t}}d $ will be executed. We don't know the exact value of $q_{m+t}$, but we can underestimate it easily  estimating the minimum effort needed for carrying a opinion of $L$ over $U$ at that moment using the actual opinion of agent 3, $\B{}{3}$. By construction, we know that the partial execution of $w$ until this moment has carried $\B{}{2}$ under $L$.  We call this underestimation $\mbox{UE}$.

 Using proposition \ref{prop:minimumEffort} we can conclude that $$ \mbox{UE} = \lceil \log_2(\frac{\B{}{3} - L}{\B{}{3} - U}) \rceil = \lceil \log_2(\frac{U + \epsilon - L}{U + \epsilon - U}) \rceil = \lceil \log_2(\frac{U + \epsilon - L}{\epsilon}) \rceil$$
 
 Because $\B{}{2} \leq L$ at the beginning of execution of  $w_{m+t}$ we can conclude that $q_{m+t} \geq \mbox{UE}$. As $q_{m+t} = \#c(w_{m+t})$ then we have $$\#c(w_{m+t}) \geq \mbox{UE}$$

 Then we have $\#c(w_{m}) \leq \mbox{OE}$ and $\mbox{UE} \leq \#c(w_{m+t})$. 

 Now we choose $\epsilon$ such that: $\mbox{OE} < \mbox{UE}$, that is, such that $$\log_2(\frac{\B{}{3}}{\B{}{3} - U}) < \log_2(\frac{U + \epsilon - L}{\epsilon})$$

that is 

 $$\frac{\B{}{3}}{\B{}{3} - U} < \frac{U + \epsilon - L}{\epsilon}$$

 then

 $$\epsilon < \frac{(U - L)(\B{}{3} - U)}{U}$$

 Then, we can conclude that $$\#c(w_{m+t}) \geq \mbox{UE} > \mbox{OE} \geq \#c(w_{m})$$ 
 \end{proof}
\fairnesshierarchy*

\begin{proof}
\label{proof:fairness-hierarchy}
We will divide the proof into the following statements to prove:
\begin{itemize}
    \item $F(k)\subset F(k+1)$: it will be divided into the following two statements to prove:
    
   \begin{itemize}
    
   \item $F(k)\subseteq F(k+1)$: If $w $ $\in$ $F(k)$ then all the $k$ windows of $w$ are complete. Then, every $e \in E$ will occur within a window of length $k$ from any position of $w$. Therefore, every $e$ $\in$ $E$ will also occur within a window of length $k+1$ (or greater) from any position of $w$, i.e., all $k+1$ windows of $w$ are complete; thus, $w$ $\in$ $F(k+1)$.  
   
   \item $F(k+1)$ $\not \subseteq$ $F(k)$: Consider a $k+1$-fair word $w = u \cdot u \cdot \ldots$  and the $k+1$-windows $u = v \cdot e_i$, such that $v(k)=E'$, where $E' \subset E$ and $E = E' \cup \{ e_i \}$, clearly every $k+1$-windows in $w$ is complete, but the occurrences of the $k$-window $v$ in $w$ are not complete. Therefore, $w \in F(k+1)$ but   $w \not \in F(k)$. 

   \end{itemize}

\item $ F(k+1) \subset (\bigcup_{\kappa\in\Nat}F(\kappa)) $: it will  be divided into the following two statements to prove:
    
   \begin{itemize}
    
   \item $ F(k+1) \subseteq (\bigcup_{\kappa\in\Nat}F(\kappa)): $ as $k+1$  $\in$  $\Nat$, it is straightforward that $ F(k+1) \subseteq (\bigcup_{\kappa\in\Nat}F(\kappa)) $.
   
   \item $  (\bigcup_{\kappa\in\Nat}F(\kappa)) \not \subseteq F(k+1) $: as $F(k+2)$ $\not \subseteq$ $F(k+1)$ and $ F(k+2) \subseteq (\bigcup_{\kappa\in\Nat}F(\kappa))$, there is at least a word $w$ such that $w \in F(k+2)$ and hence $ w \in \bigcup_{\kappa\in\Nat}F(\kappa)$ where $w \not  \in F(k+1)$.

   \end{itemize}

    \item $ (\bigcup_{\kappa\in\Nat}F(\kappa))\subset F(\omega) $: it will be divided into the following two statements to prove:
    
   \begin{itemize}
    
   \item $ (\bigcup_{\kappa\in\Nat}F(\kappa))\subseteq F(\omega) $: if $w$ $\in$ $ (\bigcup_{\kappa\in\Nat}F(\kappa))$, then $w$ $\in$ $F(k)$ for some $k \in \Nat $, therefore, every $e$ $\in$ $E$ will also occur within a window of length $k$  from any position of $w$, thus, every $e$ $\in$ $E$  occurs infinitely often in $w$, hence $w$ is strong fair; $ w \in F(\omega)$.
   
   
   \item  $ F(\omega)  \not \subseteq (\bigcup_{\kappa\in\Nat}F(\kappa))$: consider a strong-fair word $w$, i.e.  $w$ $\in $ $F(\omega)$, a set 
   $E = \{e_1, e_2\}$ where $w(i)= e_1$ if $i$ is a power of $2$, otherwise, $w(i)= e_2$; as the  distance between consecutive occurrences of  $e_1$ grows unboundedly in $w$, $e_1$ will not be  within every window of length $k$ in $w$ for some $k$ $\in$ $\Nat$. Therefore, $w$ $\not \in (\bigcup_{\kappa\in\Nat}F(\kappa))$.

   \end{itemize}
    
\end{itemize}

\end{proof}
\mfairnesshierarchy*

\begin{proof}
\label{proof:m-fairness-hierarchy}
We will divide it into the following statements to prove:
\begin{itemize}
    \item $F(k)\subset F(m+1,k)$: it will be divided into the following two statements to prove:
    
   \begin{itemize}
    
   \item $F(k)\subseteq F(m+1, k)$: If $w $ $\in$ $F(k)$ then all the $k$ windows of $w$ are complete. Therefore, from any position of $w$, every $e$ $\in$ $E$ will occur within a window of length $k$; hence, from any position of $w$ an infinite number of \emph{consecutive} complete $k$-windows appear, thus, every suffix of $w$ has a complete $(n,k)$ multi-window for all $n\in\Nat^+$, including when $n = m+1$, then $w  \in   F(m+1, k)$.

   \item $F(m+1, k)$ $\not \subseteq$ $F(k)$: Consider a  word $w$ and $E = \{e_1, e_2\}$ where $w = w_1\cdot w_2 \cdot w_3 \ldots$  such that every subword $w_{2i+1}$ corresponds  to a complete $(m+1,k)$ multi-window and every subword $w_{2i}$ corresponds to a  $k$ window of $e_1$, i.e,  $w$ can be seen as a sequence of complete $(m+1,k)$ multi-windows separated between every pair of multi-windows  by a $k$ windows of $e_1$. Clearly, $w$ $\in$ $F(m+1, k)$ as  every suffix of $w$ has a complete $(m+1,k)$ multi-window, however, as there are $k$ windows of $e_1$ in $w$, it is not true that from any position of $w$, every $e$ $\in$ $E$ will occur within a window of length $k$, therefore $w \not \in F(k)$.

   \end{itemize}

\item $F(m+1,k)\subset F(m,k)$: it will  be divided into the following two statements to prove:
    
   \begin{itemize}
    
   \item $ F(m+1,k) \subseteq F(m,k)$:  If $w $ $\in$ $F(m+1,k)$, then from any position of $w$, $m+1$ \emph{consecutive} complete $k$-windows appear, therefore, clearly from any position of $w$, $m$ \emph{consecutive} complete $k$-windows appear, i.e. $w $ $\in$ $F(m,k)$.
   
   \item $ F(m,k) \not \subseteq F(m+1,k)$: Consider a  word $w$ and $E = \{e_1, e_2\}$ where $w = w_1\cdot w_2 \cdot w_3 \ldots$  such that every subword $w_{2i+1}$ corresponds  to a complete $(m,k)$ multi-window and every subword $w_{2i}$ corresponds to a  $k$ window of $e_1$, i.e,  $w$ can be seen as a sequence of complete $(m,k)$ multi-windows separated between every pair of multi-windows by a $k$ windows of $e_1$. Clearly, $w$ $\in$ $F(m, k)$ as  every suffix of $w$ has a complete $(m,k)$ multi-window, however, there are no $m+1$ \emph{consecutive} complete $k$-windows  in  $w$, therefore $w \not \in F(m+1,k)$.
   
   \end{itemize}
    
    \item $F(m,k)\subset \bigcup_{\kappa \in \Nat} F(m,\kappa)$: it will be divided into the following two statements to prove:
    
   \begin{itemize}
    
   \item $F(m,k)\subseteq \bigcup_{\kappa \in \Nat} F(m,\kappa)$: 
    as $k$  $\in$  $\Nat$, it is straightforward that $F(m,k)\subseteq \bigcup_{\kappa \in \Nat} F(m,\kappa)$.
   
   \item  $ (\bigcup_{\kappa \in \Nat} F(m,\kappa)) \not \subseteq F(m,k) $: Consider a  word $w$ and $E = \{e_1, e_2, \ldots, e_{k+1} \}$ such that  $w = w'\cdot w' \ldots$  where $w'=e_1 \cdot e_2 \ldots e_{k+1}$. i.e. $w$ $\in$ $F(m, k+1)$. Additionally,  as $F(m,k+1)\subseteq  \bigcup_{\kappa \in \Nat} F(m,\kappa)$, $w$ $\in$ $(\bigcup_{\kappa \in \Nat} F(m,\kappa))$, however,  since $|E| = k +1 $, it is not possible that there are complete $k$-windows in $w$, therefore $w \not \in F(m,k)$.

   \end{itemize}

\item $(\bigcup_{\kappa \in \Nat} F(m,\kappa))\subset F(1,\omega)$: it will be divided into the following two statements to prove:
    
   \begin{itemize}
    
   \item $(\bigcup_{\kappa \in \Nat} F(m,\kappa))\subseteq F(1,\omega)$  : 
   if $w$ $\in$ $ (\bigcup_{\kappa\in\Nat}F(\kappa))$, then $w$ $\in$ $F(m,\kappa)$ for some $\kappa \in \Nat $, therefore, every suffix of $w$ has a complete $(m,\kappa)$ multi-window. i.e. $m$ \emph{consecutive} complete $k$-windows appear infinitely often in $w$. This implies that in every suffix of $w$, every $e$ $\in$ $E$  occurs. Therefore,  every suffix of $w$ can be seen as a complete window of infinite length ($\omega$), hence $ w \in F(1,\omega)$.

   \item  $ F(1,\omega)  \not \subseteq (\bigcup_{\kappa \in \Nat} F(m,\kappa))$: consider any $\kappa \in \Nat$,  a word $w$, a set $E = \{e_1, e_2\}$ where $w(i)= e_1$ if $i$ is a power of $2$, otherwise, $w(i)= e_2$; since $e_1$ and $e_2$ occur infinitely often in $w$,  every suffix of $w$ can be seen as a complete window of infinite length ($\omega$), hence $ w \in F(1,\omega)$.  However, since the  distance between consecutive occurrences of $e_1$ grows unboundedly in $w$, it is not possible that a complete $\kappa$-window occurs infinitely often in $w$, therefore, there are  suffixes of $w$ that don't have a complete $(m,\kappa)$ multi-window, thus,  $ w \not \in \subseteq (\bigcup_{\kappa \in \Nat} F(m,\kappa))$.

   \end{itemize}
   
\item   $F(1,\omega)=F(\omega)$:   $w$ $\in$ $F(1,\omega)$ corresponds to say that every suffix of $w$ has a complete $(1,\omega)$ multi-window, it is equivalent to say that every $e$ $\in$ $E$  occurs from any position in $w$, it equates to  say that  $w \in F(\omega)$.    
   
\end{itemize}

\end{proof}
\dynamicConsensus*

\begin{proof}
\label{proof:dynamicConsensus}
     We have $I^t_{ij} \in [\I{L},\I{U}]$ by bounded influence. We will reformulate \autoref{prop:opinionBoundAfter1step}, lemmas \ref{lemma:OpinionBoundAfterNSteps}, \ref{lemma:OpinionBoundOfAdjacentAgent}, \ref{lemma:OpinionBoundPath}, \ref{lemma:opinionSpread}, \ref{lemma:epsilonBoundedWaitESuffix}, \autoref{lemma:nepsilon} and \autoref{lemma:deltaBoundConsensus} for consensus with $\I{L}$ and $\I{U}$ playing the roles of $\I{min}$ and $\I{max}$.
     
    Similarly to \autoref{prop:opinionBoundAfter1step}, we prove that for any transition $\B{}{} \xrightarrow{(i,j)} \B[B']{}{}$ and any $k \in \agents$, we can show
    
        $$\B[B']{}{k} \leq \B{}{k}(1-\I{U}) + max(\B{}{})\I{U}$$

        Let $\pi'=\execution$ be an e-suffix of $\pi$ and $w'=e_0e_1$ the $\omega$-word generated by $\pi'$. Similarly to \autoref{lemma:OpinionBoundAfterNSteps}, we can deduce that for any $i \in \agents$, 
    
    $$\B{n}{i} \leq max(\B{0}{})-(1-\I{U})^n(max(\B{0}{})-\B{0}{i})$$
     
     Then, similar to \autoref{lemma:OpinionBoundOfAdjacentAgent}, for any $i,j \in \agents$,  if $e_{n+1} = (i,j)$, 
         $$\B{n+1}{j} \leq max(\B{0}{})-\I{L}(1-\I{U})^n(max(\B{0}{})-\B{0}{i})$$
    
    and this bound can be extended along a path, similar to \autoref{lemma:OpinionBoundPath}. Let $p$ be a path in $\graph$ that starts with agent $i$ and ends with agent $j$. Then
    
         $$\B{\delta_{w'}(p)}{j} \leq max(\B{0}{})-\I{L}^{|p|}(1-\I{U})^{\delta_{w'}(p)}(max(\B{0}{})-\B{0}{i})$$
    
    and similar to \autoref{lemma:opinionSpread}, we can show that because $\mathcal{G}$ is strongly connected, then for all $j \in \agents$ and some $i \in \agents$ 
    \begin{equation}\label{eq:opinionSpreadDynamic}
        max(\B{0}{})-\B{\Delta_{w'}(i)}{j} \geq \I{L}^{|\agents|}(1-\I{U})^{\Delta_{w'}(i)}(max(\B{0}{})-\B{0}{i})
    \end{equation}

    As proven in \autoref{lemma:mbfDeltaBound}, $\pi$ being $m$-bounded fair with $m \geq |A|-1$ implies that there exists a $\beta \in \mathbb{N}$ such that $\beta$ is a recurrent $\Delta$-bound of $\pi$. Therefore,
    
    Similarly to \autoref{lemma:epsilonBoundedWaitESuffix}, if the e-suffix $ \pi'=\B{t}{}\xrightarrow{e_t}\B{t+1}{}\xrightarrow{e_{t+1}}\dots$ of $ \pi$ has $\beta$ as a bound $\Delta$, then:

    $$ \epsilon   \leq max(\binit) - max(\B{t+\beta}{}) $$
    
    for $\epsilon = \I{L}^{|\agents|}(1-\I{U})^{\beta}(U-L)$, where  $U = \lim_{t \to \infty} max(\B{t}{})$ and $ L =  \lim_{t \to \infty} min(\B{t}{})$ from \autoref{theo:extremeLimits}. 
    
    Similarly to \autoref{lemma:nepsilon}, for all $n \in \mathbb{N}$, there exists an e-suffix $\pi'=\B{t}{}\xrightarrow{e_t}\B{t+1}{}\xrightarrow{e_{t+1}}\dots$ of $\pi$ such that $\beta$ is a $\Delta$-bound of $\pi'$ where 

    $$ n*\epsilon   \leq max(\binit) - max(\B{t+\beta}{}) $$

     for $\epsilon = \I{L}^{|\agents|}(1-\I{U})^{\beta}(U-L)$,  
    
    We are ready to reformulate \autoref{lemma:deltaBoundConsensus} for consensus with dynamic influence. It remains to prove that $U=L$.

Suppose, by contradiction, that $U \neq L$. We have $\I{U} < 1$ by bounded influence. Then $\epsilon$ is greater than or equal to a positive constant. Using \autoref{lemma:nepsilon}, take $n \in \Nat$ such that $n*\epsilon > max(\binit)$, then $max(\B{t+\beta}{}) < 0$ for some $t$, a contradiction by definition.
    
Therefore, $U=L$, and by the squeeze theorem \cite{RealAnalysis}, 
    $$\lim_{t \to \infty} max(\B{t}{}) = \lim_{t \to \infty} min(\B{t}{}) = \lim_{t \to \infty} \B{t}{k}  \;\; \forall k \in \agents $$

we converge to consensus.

\end{proof}

\extremeBounds*

\begin{proof}
\label{proof:extremeBound}
We want to prove that $$\B[B']{}{k} \leq max(\B{}{})$$

Take the transition relation of \autoref{def:ots} and let $e=(i,j)$. Either $k \neq j$ or $k = j$.

If $k \neq j$, $\B[B']{}{k} = \B{}{k}$, then $\B[B']{}{k} = \B{}{k} \leq max(\B{}{})$.

If $k = j$, 

$$\B[B']{}{k} = \B{}{j} + (\B{}{i} - \B{}{j})\I{ij}$$

Now note that by definition $\B{}{i} \leq max(\B{}{})$ and $\I{ij} \leq 1$. Then

$$\B[B']{}{k} \leq \B{}{j} + (max(\B{}{}) - \B{}{j}) = max(\B{}{})$$

as wanted. The proof that $\B[B']{}{k} \geq min(\B{}{})$ is analogous.

\end{proof}%
\opinionSpread*
\begin{proof}
\label{proof:opinionSpread}
    Take any $k \in \agents$. Because $\graph$ is strongly connected, there is a path $p$ between $i$ and $k$. Apply \autoref{lemma:OpinionBoundPath} to $p$

       $$\B{\delta_w(p)}{k} \leq max(\B{0}{})-\I{min}^{|p|}(1-\I{max})^{\delta_w(p)}(max(\B{0}{})-\B{0}{i})$$

       Now because this is for all $k \in \agents$, it follows that

           $$max(\B{\delta_w(p)}{}) \leq max(\B{0}{})-\I{min}^{|p|}(1-\I{max})^{\delta_w(p)}(max(\B{0}{})-\B{0}{i})$$

    and by \autoref{lemma:beliefExtremalBounds}, we know that $\B{\Delta_w(i)}{j} \leq max(\B{\delta_w(p)}{})$ for all $j \in \agents$.

    $$\B{\Delta_w(i)}{j} \leq max(\B{0}{})-\I{min}^{|p|}(1-\I{max})^{\Delta_w(i)}(max(\B{0}{})-\B{0}{i})$$

    Recall that by definition \ref{def:maximalWait}, $\delta_w(p) \leq \Delta_w(i)$, because $p$ starts with $i$. Then

        $$max(\B{\Delta_w(i)}{}) \leq max(\B{0}{})-\I{min}^{|p|}(1-\I{max})^{\Delta_w(i)}(max(\B{0}{})-\B{0}{i})$$

      Note that $|p| \leq |\agents|$, therefore

     $$max(\B{\Delta_w(i)}{}) \leq max(\B{0}{})-\I{min}^{|\agents|}(1-\I{max})^{\Delta_w(i)}(max(\B{0}{})-\B{0}{i})$$

     Now by substracting $max(\B{0}{})$ on both sides and multiplying by $-1$, we have
     
     $$max(\B{0}{})-max(\B{\Delta_w(i)}{}) \geq \I{min}^{|\agents|}(1-\I{max})^{\Delta_w(i)}(max(\B{0}{})-\B{0}{i})$$
    
\end{proof}%
\nepsilon*

\begin{proof}
\label{proof:nepsilon}

We proceed by induction on $n$ to prove that for all $n \in \mathbb{N}$, there exists an e-suffix $\pi'=\B{t}{}\xrightarrow{e_t}\B{t+1}{}\xrightarrow{e_{t+1}}\dots$ of $\pi$ such that $\beta$ is a $\Delta$-bound of $\pi'$ where 

    $$ n*\epsilon   \leq max(\binit) - max(\B{t+\beta}{}) $$
    
\begin{itemize}
    \item Base Case ($n =1$) 
    
     From \autoref{def:boundedWait} and as $\pi$ has a recurrent $\Delta$-bound $\beta$, there exists an e-suffix $\pi'  =\B{t}{}\xrightarrow{e_t}\B{t+1}{}\xrightarrow{e_{t+1}}\dots$ such that $\beta$ is a $\Delta$-bound of $\pi'$. 

     As $\beta$  is  a $\Delta$-bound of $\pi'$,  and  by applying \autoref{lemma:epsilonBoundedWaitRun}  on $\pi$ and $\pi'$, we have the following: 
 $$ \epsilon   \leq max(\binit) - max(\B{t+\beta}{}) $$

    with $\epsilon = \I{min}^{|\agents|}(1-\I{max})^{\beta}(U - L)$.     
     
     \item Inductive Step:

 As inductive hypothesis, we assume that there is an e-suffix $\pi'=\B{t}{}\xrightarrow{e_t}\B{t+1}{}\xrightarrow{e_{t+1}}\dots$ of $\pi$ such that $\beta$ is a $\Delta$-bound of $\pi'$ where 

    $$ n*\epsilon   \leq max(\binit) - max(\B{t+\beta}{}) $$

    with $\epsilon = \I{min}^{|\agents|}(1-\I{max})^{\beta}(U - L)$.

We need to prove that there exists an e-suffix $\pi''=\B{s}{}\xrightarrow{e_s}\B{s+1}{}\xrightarrow{e_{s+1}}\dots$ of $\pi$ such that $\beta$ is a $\Delta$-bound of $\pi''$ where 

    $$ (n+1)*\epsilon   \leq max(\binit) - max(\B{s+\beta}{}) $$

    with $\epsilon = \I{min}^{|\agents|}(1-\I{max})^{\beta}(U - L)$.

To prove it, we consider \autoref{def:boundedWait} and the fact that $\pi$ has a recurrent $\Delta$-bound  $\beta$.

As $\pi$ has a recurrent $\Delta$-bound  $\beta$, there are infinitely many e-suffixes  of $\pi$,  such that $\beta$ is a $\Delta$-bound of each of them. It implies that there are infinitely many e-suffixes of every e-suffix of $\pi$, such that  $\beta$ is a $\Delta$-bound of each of them. 

Thus, consider an e-suffix $\pi''=\B{s}{}\xrightarrow{e_s}\B{s+1}{}\xrightarrow{e_{s+1}}\dots$ of $\pi'$ such that $\beta$ is a $\Delta$-bound of $\pi''$ and $s \geq t + \beta$.

From the inductive hypothesis, we know that: 

    $$ n*\epsilon   \leq max(\binit) - max(\B{t+\beta}{}) $$


From \autoref{corollary:extremeMonotonicity} and $s \geq t + \beta$, we know that: 

$$ max(\B{t+\beta}{}) \geq max(\B{s}{}) $$

Applying \autoref{lemma:epsilonBoundedWaitESuffix}  on $\pi''$, we know that: 

$$ max(\B{s}{}) \geq max(\B{s+\beta}{})+ \epsilon $$

Then $max(\B{t+\beta}{}) \geq max(\B{s+\beta}{})+ \epsilon$ and using the inductive hypothesis, we have:

 $$ n*\epsilon   \leq max(\binit) - max(\B{t+\beta}{}) \leq max(\binit) - max(\B{s+\beta}{}) - \epsilon $$

Therefore: 

$$ n*\epsilon + \epsilon = (n + 1)*\epsilon \leq max(\binit) - max(\B{s+\beta}{})  $$

As expected. 
\end{itemize}

\end{proof}%
\deltaBoundConsensus*

\begin{proof}
\label{proof:deltaBoundConsensus}
    Let $\pi=\execution$. By \autoref{theo:extremeLimits}, there exists $U,\,L \in [0,1]$ such that $U= \lim_{t \to \infty} max(\B{t}{})$ and $L =  \lim_{t \to \infty} min(\B{t}{})$.
    
    We now prove $U=L$. Let $\beta \in \Nat$ be a recurrent $\Delta$-bound of $\pi$.

    By \autoref{lemma:nepsilon}, there exists an e-suffix $\pi'=\B{t}{}\xrightarrow{e_t}\B{t+1}{}\xrightarrow{e_{t+1}}\dots$ of $\pi$ such that $\beta$ is a $\Delta$-bound  of $\pi'$ where 

     \begin{equation}\label{eq:consensusProof1}
        n*\epsilon  \leq max(\binit) - max(\B{t+\beta}{})
    \end{equation}
    
    with $\epsilon \geq \I{min}^{|\agents|}(1-\I{max})^{\beta}(U - L)$  For all $n \in \mathbb{N}$.

    Now suppose, by contradiction, that $U \neq L$.  That, together with $\I{max} < 1$ from the puppet freedom property, imply that $\epsilon$ is a constant greater than zero. Then $U - L > 0$. Thus, from (\ref{eq:consensusProof1}) there exist some $n \in \Nat$ such that $n*\epsilon > max(\binit)$, which implies  $max(\B{t+\beta}{}) < 0$, which a contradiction by definition.
    
    Therefore, $U=L$, and by the squeeze theorem \cite{RealAnalysis}, 
    $$\lim_{t \to \infty} max(\B{t}{}) = \lim_{t \to \infty} min(\B{t}{}) = \lim_{t \to \infty} \B{t}{k}  \;\; \forall k \in \agents $$

    which is consensus.

\end{proof}%
\mbfDeltaBound*

\begin{proof}
\label{proof:mbfDeltaBound}
    Consider an OTS $\model$ with a strongly-connected $\graphElem$. Let $\pi$ be a $m$-bounded fair run with $m \geq |\agents|-1$ and let $w=w_\pi$ be its related $\omega$-word.

    If $w$ is $m$-bounded fair, it is $(m,k)$-fair for some $k$. Then, every suffix of $w$ has a complete $(m,k)$ multi-window. Let $w'=w_0\cdot w_1\dots w_m\dots$ be one of the infinitely many suffixes that start when the complete $(m,k)$ multi-window starts. Every $w_i$ is a complete $k-$window.
    
    Let $p=a_1a_2 \dots a_{|p|-1}$ be any g-path in $\graph$. 

    Because $w_1$ is complete, $a_1$ must occur in $w_1$; $a_2$ must occur in $w_2$, and so on. $a_{|p|-1}$ must occur in $w_{|p|-1}$. The path $p$ can be length $|\agents|-1$ at most, therefore every edge of $p$ must occur in order in the complete $(|\agents|-1,k)$ multi-window. This means $p$ is a subsequence of the multi-window. The size of the multi-window is greater or equal than $(|\agents|-1)*k$,  
    therefore, $\delta_{w'}(p) \leq (|\agents|-1)*k$ for any $p$. 

    This includes the g-paths that start with $m_{\pi'}$, then $\Delta_{w'}(m_{\pi'}) \leq (|\agents|-1)*k$.

    This holds for infinitely many suffixes $w'$ of $w$, specifically the suffixes that start when the complete $(m,k)$ multi-window starts.

\end{proof}%
\begin{proposition}[Minimum effort for pulling an agent over a given opinion]\label{prop:minimumEffort}
Suppose that $M=(\graph,\binit,\to)$ where $\graph$ is the strongly-connected, puppet-free, influence graph in Fig. \ref{fig:ex1} and $\binit=(0,0.5,1)$ as in example \ref{ex:edgeFairnessCounterEx}.
Let $i,j \in \agents, (j,i)\in E$ and $U \in [0,1]$ such that $\B{}{i} < U < \B{}{j}$. Let $t \in \Nat$, the least number of consecutive executions of edge $(j,i)$ needed for pulling the opinion of agent $i$ over $U$. Then  $t = \lceil \log_2(\frac{\B{}{j} - \B{}{i}}{\B{}{j} - U}) \rceil$.
\end{proposition}
\begin{proof}
\label{proof:minimumEffort}
    Let $\delta = \B{}{j} - \B{}{i}$. The opinion of agent $i$ after $t$ consecutive activations of edge $(j,i)$ is  $\B{}{i} + (\frac{\delta}{2^1} + \frac{\delta}{2^2} + \ldots + \frac{\delta}{2^T})$. We need to find $t$ such that this value be over $U$. That is: $$ \B{}{i} + (\frac{\delta}{2^1} + \frac{\delta}{2^2} + \ldots + \frac{\delta}{2^t}) \geq U $$

    We know that $ \B{}{i} + (\frac{\delta}{2^1} + \frac{\delta}{2^2} + \ldots + \frac{\delta}{2^T}) = \B{}{i} + \delta(1- \frac{1}{2^t})$.

    Let $T$ be the value where $\B{}{i} + \delta(1- \frac{1}{2^T}) = U $, that is, solving $T$: $$T=\log_2(\frac{\B{}{j} - \B{}{i}}{\B{}{j} - U})$$

    Then $t=\lceil T \rceil $
\end{proof}
\begin{lemma}[Opinion bound after one step]\label{prop:opinionBoundAfter1step}
    For any transition $\B{}{} \xrightarrow{\cdot} \B[B']{}{}$, and any $k \in \agents$

    $$\B[B']{}{k} \leq \B{}{k}(1-\I{max}) + max(\B{}{})\I{max}$$
\end{lemma}

\begin{proof}
\label{proof:greaterStep}
Take the transition relation of \autoref{def:ots}, where
$k \neq j$ implies $\B[B']{}{k} = \B{}{k}$ and $k = j$ implies $\B[B']{}{k} =  \B{}{j} + (\B{}{i} - \B{}{j})\I{ij}$. For the first case $\B[B']{}{k} = \B{}{k}$,

\begin{align*}
\B[B']{}{k} &\leq \B{}{k}\\
\B[B']{}{k} &\leq \B{}{k}(1 - \I{max} + \I{max})\\
\B[B']{}{k} &\leq \B{}{k}(1 - \I{max}) + \B{}{k}\I{max}\\
\B[B']{}{k} &\leq \B{}{k}(1 - \I{max}) + max(\B{}{})\I{max}&~ &\mbox{ by definition } \B{}{k} \leq max(\B{}{})\\
\end{align*}

For the second case, we have

$$\B[B']{}{k} =B{}{j} + (\B{}{i} - \B{}{j})\I{ij}$$

by definition, $\B{}{i} \leq max(\B{}{})$ and $\I{ij} > 0$, therefore

$$\B[B']{}{k} \leq \B{}{j} + (max(\B{}{})-\B{}{j})\I{ij}$$

In addition, $\I{ij} \leq \I{max}$ by definition; replacing

$$\B[B']{}{k} \leq \B{}{j} + (max(\B{}{})-\B{}{j})\I{max}$$ 

which can be rewritten as

$$\B[B']{}{k} \leq \B{}{k}(1-\I{max}) + max(\B{}{})\I{max}$$

\end{proof}%
\begin{restatable}[Opinion upper bound after $n$ steps]{lemma}{OpinionBoundAfterNSteps}\label{lemma:OpinionBoundAfterNSteps}
    Let $\execution$ be an e-path.
    For any $i \in \agents$
    $$\B{n}{i} \leq max(\B{0}{})-(1-\I{max})^n(max(\B{0}{})-\B{0}{i})$$
\end{restatable}

\begin{proof}
\label{proof:OpinionBoundAfterNSteps}



We use induction for $n$. For the base case $n=1$, we claim

$$\B{1}{i} \leq max(\B{}{}) - (1-\I{max})(max(\B{}{}) - \B{}{i})$$

which we deduce by applying \autoref{prop:opinionBoundAfter1step} to $\B{}{} \xrightarrow{e_1} \B{1}{}$:

\begin{align*}
    \B{1}{i} &\leq \B{}{i}(1-\I{max}) + max(\B{}{})(\I{max})\\
    &= \B{}{i}- \B{}{i}(\I{max}) + max(\B{}{})(\I{max})\\
    &= \B{}{i}+ \I{max}(max(\B{}{})-\B{}{i})\\
    &= max(\B{}{}) - max(\B{}{}) + \B{}{i}  +\I{max}(max(\B{}{})-\B{}{i})\\
    &= max(\B{}{}) - (max(\B{}{}) - \B{}{i})  +\I{max}(max(\B{}{})-\B{}{i})\\
    &= max(\B{}{}) - (1-\I{max})(max(\B{}{})- \B{}{i})\\
\end{align*}

as claimed. For the inductive case, we claim

\begin{multline}
    \B{n}{i} \leq max(\B{}{}) - (1-\I{max})^n(max(\B{}{})- \B{}{i})  \implies \\ 
    \B{n+1}{i} \leq max(\B{}{}) - (1-\I{max})^{n+1}(max(\B{}{})- \B{}{i})
\end{multline}

to prove this, we apply \autoref{prop:opinionBoundAfter1step} to $\B{n}{} \xrightarrow{e_{n+1}} \B{n+1}{}$:

$$\B{n+1}{i} \leq \B{n}{i}(1-\I{max}) + max(\B{n}{})\I{max}$$

by \autoref{lemma:beliefExtremalBounds}, $max(\B{n}{}) \leq max(\B{}{})$

$$\B{n+1}{i} \leq \B{n}{i}(1-\I{max}) + max(\B{}{})\I{max}$$

using the inductive hypothesis, we replace $\B{n}{i}$

\begin{align*}
    \B{n+1}{i} &\leq \left[max(\B{}{}) - (1-\I{max})^n(max(\B{}{})- \B{}{i})\right](1-\I{max}) + max(\B{}{})\I{max}\\
    &= max(\B{}{}) -  \left[(1-\I{max})^n(max(\B{}{})- \B{}{i})\right](1-\I{max})\\
    &= max(\B{}{}) - (1-\I{max})^{n+1}(max(\B{}{})- \B{}{i})\\
\end{align*}
as claimed. Thus, for any $n \in \mathbb{N}$ and $i \in \agents$,

$$\B{n}{i} \leq max(\B{}{}) - (1-\I{max})^n(max(\B{}{})- \B{}{i}) $$

\end{proof}%
\begin{restatable}[Direct influence bound]{lemma}{OpinionBoundOfAdjacentAgent}\label{lemma:OpinionBoundOfAdjacentAgent}
    Consider the e-path $\execution$ with $e_{n+1}=(i,j)$. Then 
    $$\B{n+1}{j} \leq max(\B{}{})-\I{min}(1-\I{max})^n(max(\B{}{})-\B{}{i})$$
\end{restatable}

\begin{proof}
\label{proof:OpinionBoundOfAdjacentAgent}
From $e_{n+1}=(i,j)$, we apply the $\alpha$ update function of \autoref{def:ots} to $\B{n}{j}$ once to get

\begin{equation}\label{eq1:OpinionBoundOfAdjacentAgent}
\alpha(\B{n}{i},\B{n}{j}, \I{ij}) = \B{n+1}{j} = \B{n}{j}(1-\I{ij}) + \B{n}{i}\I{ij}
\end{equation}

by \autoref{lemma:OpinionBoundAfterNSteps}, we know that

$$\B{n}{i} \leq max(\B{}{})-(1-\I{max})^n(max(\B{}{})-\B{}{i})$$

replacing $\B{n}{i}$ in (\ref{eq1:OpinionBoundOfAdjacentAgent})

\begin{equation}\label{eq2:OpinionBoundOfAdjacentAgent}
\B{n+1}{j} \leq \B{n}{j}(1-\I{ij}) + max(\B{}{})\I{ij} - (1-\I{max})^n(max(\B{}{})-\B{}{i})\I{ij}
\end{equation}

and because $\B{n}{j} \leq max(\B{}{})$ from \autoref{lemma:beliefExtremalBounds}, we infer 

$$\B{n}{j}(1-\I{ij}) + max(\B{}{})\I{ij} \leq max(\B{}{})$$

therefore replacing on (\ref{eq2:OpinionBoundOfAdjacentAgent}) yields

$$ \B{n+1}{j} \leq max(\B{}{})-\I{ij}(1-\I{max})^n(max(\B{}{})-\B{}{i})$$

By definition, $\I{min} \leq \I{ij}$, therefore

$$\B{n+1}{j} \leq max(\B{}{})-\I{min}(1-\I{max})^n(max(\B{}{})-\B{}{i})$$

\end{proof}%
%

\begin{restatable}[Opinion upper bound along a path]{lemma}{OpinionBoundPath}\label{lemma:OpinionBoundPath}
Let $\pi=\execution$ be an e-path such that $w_\pi$ is strongly fair. Let $p$ be a g-path in $\graph$ from agent $i$ to $j$. Then

     $$\B{\delta_w(p)}{j} \leq max(\B{0}{})-\I{min}^{|p|}(1-\I{max})^{\delta_w(p)}(max(\B{0}{})-\B{0}{i})$$
\end{restatable}

\begin{proof}
\label{proof:OpinionBoundPath}
    Let $p = p_0\dots p_k \dots p_{|p|-1};\; p_k=(i_k, i_{k+1})$. Define 
    
    $$\tau(k) = \delta_w(p_0\dots p_k)$$ 
    
    Notice that $\tau(k) \in \mathbb{N}$ because $\delta_r(p) \in \mathbb{N}$. $\tau(k)$ is time step when the $k$-th edge of $p$ is activated after $\tau(k-1)$, i.e. $e_{\tau(k)} = p_k$. 

    We assert that
    \begin{equation}\label{proof:opinionBoundPathClaim}
        \B{\tau(k)}{i_{k+1}} \leq max(\B{0}{})-\I{min}^{k+1}(1-\I{max})^{\tau(k)}(max(\B{0}{})-\B{}{i})
    \end{equation}

     proving it by induction over $k$. For the base case $k = 0$, we claim 

   \begin{equation}
        \B{\tau(0)}{i_{1}} \leq max(\B{0}{})-\I{min}(1-\I{max})^{\tau(0)}(max(\B{0}{})-\B{}{i})
    \end{equation}

   Consider the e-path $\B{0}{} \to \B{1}{}\to \B{2}{} \dots \xrightarrow{p_0} \B{\tau(0)}{}$. Applying \autoref{lemma:OpinionBoundOfAdjacentAgent}:

    \begin{equation}
        \B{\tau(0)}{i_{1}} \leq max(\B{0}{})-\I{min}(1-\I{max})^{\tau({0})-1}(max(\B{0}{})-\B{}{i})
   \end{equation}

   and we can prove that $(1-\I{max})^{\tau({0})-1} \geq (1-\I{max})^{\tau({0})}$, therefore we derive

       \begin{equation}
        \B{\tau(0)}{i_{1}} \leq max(\B{0}{})-\I{min}(1-\I{max})^{\tau({0})}(max(\B{0}{})-\B{}{i})
   \end{equation}

   as claimed. For the inductive case, we need to prove

   \begin{multline*}
       \B{\tau(k)}{i_{k+1}} \leq max(\B{0}{})-\I{min}^{k+1}(1-\I{max})^{\tau(k)}(max(\B{0}{})-\B{}{i}) \implies \\
       \B{\tau(k+1)}{i_{k+2}} \leq max(\B{0}{})-\I{min}^{k+2}(1-\I{max})^{\tau(k+1)}(max(\B{0}{})-\B{}{i})
   \end{multline*}

    Consider the e-path $\B{\tau({k})}{} \to \B{\tau({k})+1}{} \to \B{\tau({k})+2}{} \dots \xrightarrow{p_{k+1}} \B{\tau(k+1)}{}$ which is an e-suffix of $\pi$. From \autoref{lemma:OpinionBoundOfAdjacentAgent} we obtain:

    \begin{equation}
        \B{\tau(k+1)}{i_{k+2}} \leq max(\B{0}{})-\I{min}(1-\I{max})^{\tau({k+1})-\tau({k})-1}(max(\B{0}{})-\B{\tau({k})}{i_{k+1}})
    \end{equation}

    replacing $\B{\tau({k})}{i_{k+1}}$ given the inductive hypothesis

    \begin{multline*}
        \B{\tau(k+1)}{i_{k+2}} \leq max(\B{0}{})-\I{min}(1-\I{max})^{\tau({k+1})-\tau({k})-1}(max(\B{0}{})-\\
        \left[max(\B{0}{})-\I{min}^{k+1}(1-\I{max})^{\tau(k)}(max(\B{0}{})-\B{}{i}) \right])\\
        = max(\B{0}{})-\I{min}^{k+2}(1-\I{max})^{\tau({k+1})-1}(max(\B{0}{})-\B{}{i})
    \end{multline*}

    and we can prove that $(1-\I{max})^{\tau(k+1)-1} \geq (1-\I{max})^{\tau(t+1)}$, therefore we derive

    $$\B{\tau(k+1)}{i_{k+2}} \leq max(\B{0}{})-\I{min}^{k+2}(1-\I{max})^{\tau(k+1)}(max(\B{0}{})-\B{}{i})$$
    
    as wanted. Thus, (\ref{proof:opinionBoundPathClaim}) is true. 
    Then, for $k=|p|-1$,

    \begin{equation*}
        \B{\tau(|p|-1)}{i_{|p|}} \leq max(\B{0}{})-\I{min}^{|p|}(1-\I{max})^{\tau(|p|-1)}(max(\B{0}{})-\B{}{i})
    \end{equation*}

    clearly, $i_{|p|}=j$ and $\tau(|p|-1) = \delta_w(p)$, therefore

        \begin{equation*}
        \B{\delta_w(p)}{k} \leq max(\B{0}{})-\I{min}^{|p|}(1-\I{max})^{\delta_w(p)}(max(\B{0}{})-\B{}{i})
    \end{equation*}

\end{proof}%
\begin{restatable}[L and U are bounds of extremes]{corollary}{LUareBounds}\label{corollary:LUareBounds}
     Let $\execution$ be an e-path and $U,\,L \in [0,1]$ such that $\lim_{t \to \infty} \{max(\B{t}{})\} = U$ and $\lim_{t \to \infty} \{min(\B{t}{})\} = L$. Then $max(\B{t}{}) \geq U$ and $min(\B{t}{}) \leq L$  for all $t \in \mathbb{N}$.
 \end{restatable}




\begin{restatable}[epsilon decrement $\Delta$-bound e-suffix]{lemma}{epsilonDecrementBoundedWaitE-Suffix}\label{lemma:epsilonBoundedWaitESuffix}

Let $\pi'=\B{0}{}\xrightarrow{e_0}\B{1}{}\xrightarrow{e_1}\dots$ be an e-suffix of a strongly-fair run, with a $\Delta$-bound $\beta \in \mathbb{N}$, then:


    $$  max(\B{0}{}) - max(\B{\beta}{})\geq \epsilon $$

    with $\epsilon = \I{min}^{|\agents|}(1-\I{max})^{\beta}(U - L)$.

\end{restatable}

\begin{proof}
\label{proof:epsilonDecrement}
Considering $\pi'$ and the agent $ m_{\pi'} \in \agents$,  we apply  \autoref{lemma:opinionSpread}  to bound the maximum opinion of an agent in $\pi'$ as follows: 
$$max(\B{0}{}) - max(\B{\Delta_{w_\pi'}(m_{\pi'})}{})  \geq \I{min}^{|\agents|}(1-\I{max})^{\Delta_{w_\pi'}(m_{\pi'})}(max(\B{0}{})-\B{0}{m_{\pi'}})$$

As $max(\B{0}{})\geq U $  and $\B{0}{m_{\pi'}}\leq L$ from \autoref{corollary:LUareBounds}, then: 

$$max(\B{0}{}) - max(\B{\Delta_{w_\pi'}(m_{\pi'})}{})  \geq \I{min}^{|\agents|}(1-\I{max})^{\Delta_{w_\pi'}(m_{\pi'})}(U - L)$$

 As $\beta$ is a $\Delta$-bound of $\pi'$, $\Delta_{w_\pi'}(m_{\pi'}){}$ $ \leq \beta $, hence: 

$$max(\B{0}{}) - max(\B{\Delta_{w_\pi'}(m_{\pi'})}{})\geq \I{min}^{|\agents|}(1-\I{max})^{\Delta_{w_\pi'}(m_{\pi'})}(U - L)\geq \I{min}^{|\agents|}(1-\I{max})^{\beta}(U - L)$$

From \autoref{corollary:extremeMonotonicity} and $\Delta_{w_\pi'}(m_{\pi'}){}$ $ \leq \beta $, $max(\B{\Delta_{w_\pi'}(m_{\pi'})}{}{})\geq max(\B{\beta}{})$, consequently: 

$$  max(\B{0}{}) - max(\B{\beta}{})\geq max(\B{0}{}) - max(\B{\Delta_{w_\pi'}(m_{\pi'})}{}) \geq  \epsilon $$

    with $\epsilon = \I{min}^{|\agents|}(1-\I{max})^{\beta}(U - L)$. , as expected.

\end{proof}

\begin{restatable}[epsilon decrement $\Delta$-bound run]{lemma}{epsilonDecrementBoundedWaitRun}\label{lemma:epsilonBoundedWaitRun}
\label{proof:epsilonDecrementBoundedWaitRun}
Let $\pi=\binit\xrightarrow{e_0}\B{1}{}\xrightarrow{e_1}\dots$ be a run with strongly-connected $\graph$, and an e-suffix $ \pi'=\B{t}{}\xrightarrow{e_t}\B{t+1}{}\xrightarrow{e_{t+1}}\dots$ of $ \pi$ with a $\Delta$ bound $\beta \in \mathbb{N}$, then:

    $$  max(\binit) - max(\B{t+\beta}{})\geq \epsilon   $$

    with $\epsilon = \I{min}^{|\agents|}(1-\I{max})^{\beta}(U - L)$.

\end{restatable}

\begin{proof}

Applying \autoref{lemma:epsilonBoundedWaitESuffix} on $\pi'$, we obtain: 
$$  max(\B{t}{}) - max(\B{t + \beta}{})\geq \epsilon $$

    with $\epsilon = \I{min}^{|\agents|}(1-\I{max})^{\beta}(U - L)$

From \autoref{corollary:extremeMonotonicity}, we know $max(\binit)\geq max(\B{t}{})$, therefore:

$$max(\binit) - max(\B{t+\beta)}{} \geq max(\B{t}{}) - max(\B{t+\beta)}{} \geq \I{min}^{|\agents|}(1-\I{max})^{\beta}(U - L)$$

Then: 

 $$  max(\binit) - max(\B{t+\beta}{})\geq \epsilon   $$

    with $\epsilon = \I{min}^{|\agents|}(1-\I{max})^{\beta}(U - L)$, as expected.

\end{proof}%

\end{document}